\documentclass[12pt]{article}
\begin{document}
\input epsf
\def\be{\begin{equation}}
\def\bea{\begin{eqnarray}}
\def\ee{\end{equation}}
\def\eea{\end{eqnarray}}
\def\d{\partial}

\begin{flushright}
OHSTPY-HEP-T-00-010\\
hep-th/0006196
\end{flushright}
\vspace{15mm}
\begin{center}
{\LARGE Correlation functions for $M^N/S_N$ orbifolds}
\\
\vspace{20mm}
{\bf  Oleg Lunin  and  Samir D. Mathur \\}
\vspace{4mm}
Department of Physics,\\ The Ohio State University,\\ Columbus, OH 43210, USA\\
\vspace{4mm}
\end{center}
\vspace{5mm}
\begin{abstract}

We develop a method for computing  correlation functions of twist
operators in the bosonic 2-d CFT
arising from orbifolds $M^N/S_N$, where $M$ is  an arbitrary
manifold. The path integral with twist operators is
replaced by a path integral on a covering space with no operator
insertions.  Thus, even though the CFT is
defined on the sphere, the correlators are expressed in terms of
partition functions on  Riemann
surfaces with a finite range of genus $g$.  For large N, this genus
expansion coincides with a 1/N expansion.
The contribution from the  covering space of genus zero is
`universal' in the sense that it depends only on the
central charge of the CFT.  For 3-point functions we give an explicit
form for the contribution from the sphere, and
for the 4-point function we do an example which has genus zero and genus one
contributions. The condition for the genus zero
contribution to the 3-point functions to be non--vanishing is similar
to the fusion rules for an SU(2) WZW model.  We
observe that the 3-point coupling becomes small compared to its large
N limit when the orders of the twist
operators  become comparable to the square root of N - this is a
manifestation of the stringy exclusion principle.

\end{abstract}
\newpage

\section{Introduction}
\renewcommand{\theequation}{1.\arabic{equation}}
\setcounter{equation}{0}

The AdS/CFT correspondence gives a remarkable
relation between string theory on   a spacetime  and a certain
conformal field theory (CFT) on the
boundary of this spacetime \cite{mal}.  In particular the near horizon geometry
of D3 branes gives the space
$AdS_5\times S^5$, and string theory on this space is conjectured to
be dual to N=4 supersymmetric Yang-Mills on the
boundary of the $AdS_5$. When the string theory is weakly coupled,
tree level supergravity is a valid low energy
approximation. The dual  CFT at this point is a strongly coupled
Yang-Mills theory, which cannot therefore be studied
perturbatively. On the other hand weakly coupled Yang-Mills theory is
dual to string theory in a domain of parameters
where the latter cannot be approximated by supergravity on a gently
curved spacetime. In spite of this fact, it turns out
that certain quantities computed in the supergravity limit of string
theory agree with their corresponding dual quantities
in the Yang-Mills theory, where the latter computation is done at {\it
weak} coupling. One believes that such an agreement is
due to the supersymmetry which is present in the theory; this
supersymmetry would for example protect the dimensions
of chiral operators from changing when the coupling is varied.
Interestingly, the values of 3-point correlation functions of
chiral operators are also found to agree, when we compare the tree
level supergravity calculation on AdS space with the
computation in the free Yang-Mills theory \cite{threepoint}; the
latter is just the
result obtained by Wick contractions among the fields in
the chiral operators. It is not clear if the 3-point function of
chiral operators is protected against change of coupling at all
values of $N$; the above result just tells us that the large $N$
results agree between the  weak and strong coupling
limits.\footnote{See however \cite{howe} for an analysis of
the finite $N$ case.}

One is thus led to ask: are the 3-point functions of chiral operators
protected in the other cases of the AdS/CFT
correspondence? In particular we will be interested in the case of
the D1-D5 system \cite{mal,done}, which gives a near-horizon geometry
$AdS_3\times S^3\times M$, where $M$ is a torus $T^4$ or a  $K3$
space.  This system is of great interest for the issues
related to black holes, since it yields, upon addition of momentum
excitations,  a supersymmetric configuration which has a
classical  (i.e. not Planck size) horizon.  In particular, the
Bekenstein entropy computed from the classical horizon area
agrees with the count of microstates for the extremal and near
extremal black holes \cite{stromvafa}. Further,   the low energy  Hawking
radiation from the  hole can be understood in terms of a unitary
microscopic process, not only qualitatively but also
quantitatively, since one finds an agreement of spin dependence and
radiation rates between the semiclassically computed
radiation and the microscopic calculation \cite{dasmathur}. While it
is possible to
use simple models for the low energy dynamics of the
D1-D5 system when one is computing the coupling to massless modes of
the supergravity theory, it is believed that the
exact description of this CFT must be in terms of  a sigma model with
target space being a deformation of the orbifold
$M^N/S_N$, which is the symmetric orbifold of $N$ copies of
$M$.  (Here $N=n_1n_5$, with $n_1$ being the number of
D1 branes and $n_5$ being the number of D5 branes, and we must take
the low energy limit of the sigma model to obtain
the desired CFT.) In particular we may consider the `orbifold point'
where the target space is exactly the orbifold
$M^N/S_N$ with no deformation. It was argued in
\cite{martinec} that this CFT does correspond to a certain point in
the moduli space of string theories on $AdS_3\times
S^3\times M$, but at this point the string theory is in a strongly
coupled domain where it cannot be approximated by
tree level  supergravity on a smooth background. The orbifold point
is the closest we can get to a `free' theory on the CFT
side, and thus this point is the analogue of free N=4 supersymmetric
Yang-Mills in the D3 brane example. Thus one would
like to compare the three point functions of chiral operators in the
supergravity limit with the 3-point functions at the
orbifold point, to see if we have an analogue of the surprising
agreement that was found in the case of the
$AdS_5$ - 4-d Yang-Mills duality.

The orbifold group in our case is $S_N$, the permutation group of $N$
elements. This group is nonabelian, in contrast to the
cyclic group $Z_N$ which has been studied more extensively in the
past for computation of correlation functions in orbifold
theories \cite{dixon}. Though there are some results in the literature for
general orbifolds \cite{vafaorb}, the study of nonabelian orbifolds
is much less
developed than for abelian orbifolds.

It turns out however that the
case of the $S_N$ orbifolds has its own set of
simplifications which make it possible to develop a technique for
computation of correlation functions for these theories.  The essential
quantities that we wish to compute are the correlation functions of `twist
operators',  in the  CFT that arises from the infra-red
limit of the 2-d sigma model with target space $M^N/S_N$.   If we circle the
insertion of a twist operator of the permutation group
$S_N$,  different copies of the target space
$M$ permute into each other.  We pass to the covering space of the 2-d base
space, such that on this covering space the fields  of the CFT are
single-valued.  For the special case where the orbifold group is $S_N$, the
path integral on the base space with twist operators inserted becomes a path
integral over the covering space for the CFT with only one copy of $M$, with
no operator insertions. Thus the correlation functions of twist operators can
be rewritten as partition functions on  Riemann surfaces of different genus for
the CFT arising from one copy of $M$.

In the simplest case, which is also the case giving the leading contribution at
large $N$, the genus of the covering surface is zero, and we just get the
partition function on the sphere. But the metric on this sphere is 
determined by
the orders and locations of the twist operators. We can write this metric as
$g=e^\phi \hat g$, for some fiducial metric $\hat g$, provided we take into
account the conformal anomaly given by the Liouville action for $\phi$. It
turns out that $\phi$ is harmonic outside a finite number of isolated
points, so
the Liouville action can be computed by observing the local behavior of the
covering surface at these points.  In this manner we can compute any
correlation function of twist operators.  For given operators we get
contributions from only a finite range of genera for the covering surface.

We compute the 2-point function of twist operators, from which we recover
their well known scaling dimensions.  We then compute the
contribution to  3-point functions which comes from covering surfaces of
genus zero.  This gives the complete result
for the fusion coefficients of twist operators  for a subset of cases
(the `single
overlap'  cases) and the leading result at large $N$ for all cases. We then
compute the  4-point function for twist operators of order $2$; this
correlator has contributions from genus 0 and genus 1, and we compute
both contributions.

We find a certain `universality' in the correlation functions, since the
Liouville action depends only on the central charge $c$ of the CFT with target
space $M$. The contribution from higher genus covering surfaces will involve
the partition functions at those genera, while the leading order result coming
from the covering surface of genus zero will depend only on $c$. These
observations generalize the fact that the dimensions of the twist
operators depend only on $c$.

    We address only bosonic operators in this paper, though we
expect the extension to the supersymmetric case to be
relatively straightforward.  Thus we
also do not address  the comparison to  supergravity; this comparison
should be carried out only for the supersymmetric case. But we do observe
some features of our  correlation functions that accord with some of 
the patterns that
emerge in the supergravity computation.  In particular we find a similarity
between the condition for the 3-point correlator to have a contribution from
genus zero covering surfaces, and the condition for primaries to fuse
together in the SU(2) Wess-Zumino-Witten model.

There are several earlier works that relate to the problem we are studying, in
particular \cite{vafaone,  frolov, bantay, jevicki, mihailescu, halpern}. We will
mention these in more detail in
the context where they appear.

The plan of this paper is the following. Section 2 describes our method of
computing correlation functions.  In section 3  we compute the 
2-point functions,
and thus recovers the scaling dimensions of the twist operators.  In section 4
we construct the map to the covering space for 3-point functions, for
the case where the covering space is a sphere.  In section 5 we find the
Liouville action associated to this map. In section 6 we use the above results
to obtain the contribution to 3-point functions from covering surfaces of
genus zero. Section 7 discusses an example of a 4-point function.
Section 8 is a
discussion.

\section{Computing correlation functions through the Liouville action}
\renewcommand{\theequation}{2.\arabic{equation}}
\setcounter{equation}{0}
\subsection{The twist operators}

Let us consider for simplicity the case $M=R$, i.e. the noncompact
real line.  Then the sigma model target space
$M^N/S_N$ can be described through a collection of $N$ free bosons
$X^1, X^2,
\dots X^N$, living on a plane parameterized by the complex
coordinate $z$.  We will see later that we can extend the analysis
directly  to other CFTs. Let the $z$ plane have the flat
metric
\be
ds^2=dzd\bar z.
\label{one}
\ee
The CFT is defined through a path integral over the values of the
$X^i$, with action
\be
S=\int d^2z ~ 2\partial_z X^i\partial_{\bar z}X^i.
\label{two}
\ee
We now make the definition  of this partition function more precise.
We cut off the $z$ plane at a large radius
\be
|z|={1\over \delta}, ~~~\delta ~~{\rm small}.
\label{three}
\ee
We want boundary conditions at this large circle to represent the
fact that the identity operator has been inserted at
infinity. We will explain the norm of this boundary state later on.

Thus we imagine that our CFT is defined on a `sample' of size
$1/\delta$ --  correlation functions are to be computed by putting
the operators
at $|z_i|<<1/\delta$ and  we take $\delta$ to zero at the end of
the calculation.  (An exception will be the insertion of an operator
at infinity, which we will have to define
separately.) We write the path integral for a single boson on the $z$ plane as
\be
Z_\delta=\int DX~ e^{-S}.
\label{four}
\ee
The path integral for $N$ bosons, with no twist operator insertions,
is $(Z_\delta)^N$.

     The twist operator
$\sigma_{12}(z_1)$ can be described through the following.  Cut a
circular hole of radius
$\epsilon$ in the
$z$ plane about the point $z_1$. While the path integral over $X^i,
i=3,\dots N$ is left unchanged, we modify the boundary
conditions on $X^1, X^2$ such that as we go around the hole at $z_1$ we get
\be
X^1\rightarrow X^2, ~~X^2\rightarrow X^1.
\label{five}
\ee
We call this operator $\sigma^\epsilon_{12}(z_1)$, where the $\epsilon$ in
the superscript reminds us of the regulation used to
define the twist.  Note that we still have to define more precisely
the state that is inserted at the edge of the hole
of radius $\epsilon$ - we will do this below. If we want to maintain
the boundary condition at the circle
$|z|=1/\delta$ (and introduce no twist there) we must insert at
some point
$z_2$ another such twist operator $\sigma_{12}$; the size of the
circular hole around $z_2$ is also $\epsilon$. Let us
compute the path integral with these boundary conditions, and call it
$Z_{\epsilon, \delta}[\sigma_2(z_1), \sigma_2(z_2)]$.
Then we define the correlation function
\be
\langle\sigma^\epsilon(z_1)\sigma^\epsilon(z_2)\rangle_\delta\equiv
{Z_{\epsilon,
\delta}[\sigma_2(z_1), \sigma_2(z_2)]\over (Z_\delta)^N}.
\label{six}
\ee
We will later define rescaled twist operators, and the cutoffs
$\epsilon, \delta$ will disappear from all final answers.

\subsection{Path integral on the covering space}

The functions $X^1, X^2$ above were not single valued in the $z$
plane due to the insertion of the twist operators. We wish
to pass to a covering space where these functions would become one
single valued function. Since the fields $X^3, \dots X^N$
are not involved in the twist, they cancel out in the RHS of
(\ref{six}).  We thus consider only  $X^1, X^2$ in the
following.

Consider a configuration of the fields $X^1, X^2$ which contributes
to the path integral. Consider a simply connected patch
of the
$z$ plane, which excludes the holes around
$z_1, z_2$. Over this patch there are two functions defined: one for
each field, though due to the twists there is no global
way to label the functions uniquely as being $X^1$ or $X^2$. Take any
one of the functions over this patch - call it $X(z)$.  To
construct the covering surface $\Sigma$, let this open set in $z$ be
a patch on $\Sigma$, with the complex structure given
by $z$, and the metric also equal to the metric (\ref{one}) of the
$z$ plane. There is one field $X$ that will be defined over
$\Sigma$, and over this patch let it be the above mentioned function
$X(z)$.  Now consider another such simply connected
open patch, partly overlapping with the first, and use it to define
another patch on $\Sigma$.  Clearly, as we go around the
point $z_1$, following these overlapping patches, the surface
$\Sigma$ will look locally like the Riemann surface of a
function
\be
t=(z-z_1)^{1/2}.
\label{sixp}
\ee
     Further, the functions $X^1, X^2$ will be
both encoded in the function $X$ which will be single
valued on $\Sigma$, and the action of the configuration of $X^1, X^2$
will be reproduced if in each patch we use for $X$ the
action
\be
S=\int_{\rm patch} d^2z ~ 2\partial_z X \partial_{\bar z} X.
\label{seven}
\ee

The coordinate $z$ cannot be a globally well defined coordinate for
$\Sigma$, since a generic value of $z$ corresponds to two
points on $\Sigma$. We will call our choice of coordinate on $\Sigma$
as $t$, which will be locally holomorphically related to
$z$.

     In (\ref{seven}) we must evaluate the path integral on the patch
using the metric induced from the $z$ plane on
the patch;
the path integral depends in general on the metric and the physical
problem is defined through the metric
chosen on the $z$  plane. Thus in terms
of the coordinate $t$ used to describe $\Sigma$ we
will have
\be
ds^2=dzd\bar z = |{dz\over dt}|^2 dtd\bar t.
\label{el}
\ee

     For the example above we can take
\be
z=a{t^2\over 2t-1}.
\label{eight}
\ee
Near the location $z=0, t=0$, we evidently have $z\sim t^2$, so
that $t$ parameterizes the covering surface near $z=0$.
But the map between $z$ and $t$ is singular also at $z=a, t=1$, since
near $t=1$
\be
{dz\over dt}\approx 2a(t-1) , ~~~~ z-a\approx a(t-1)^2.
\label{nine}
\ee
Thus the twist operators are located at
$z_1=0$ and $z_2=a$.

The covering space $\Sigma$ (parameterized by $t$) is a double cover
of the original sphere parameterized by $z$.
We had cut the $z$ plane at  $|z|=1/\delta$,  and inserted the identity  there.
This boundary in the $z$ space corresponds to two regions in the $t$
space: $z\rightarrow\infty$ maps to
$t\rightarrow \infty$  as well as $t\rightarrow 1/2$. Thus in the $t$ plane we
will have a  boundary near infinity, as well as  in a small disc cut out around
$t=1/2$.  We will have the identity inserted at each of these
boundaries; the precise state with norm will be
defined in the subsection below.

We now note that we have   simply a path integral over
a single free boson $X$ on $\Sigma$  - there is no
twist operator left in the problem, and any boundaries present on
$\Sigma$ carry only the identity state.
We will show below how to close these holes in $\Sigma$, and then we
will have just a path integral over a
closed surface to compute.

The method of passing to a covering space to analyze orbifold correlation
functions has been studied by many authors, for example \cite{vafaone, dixon}. 
The observation that for symmetric orbifolds one gets a single copy of the
target space with  no nontrivial
operator insertions on the covering space  is
implicit in
\cite{frolov}. The notion of passing to the covering space to take
into account the twist operators for symmetric orbifolds is also used in the
computation of partition functions in \cite{bantay}. The $Z_2$
orbifold is the same as the $S_2$ orbifold, and the map to the
covering space was used in \cite{vafaone,dixon}  to find the 4-point
correlation of twist operators for the $Z_2$ orbifold of a complex boson.

We will depart from the usual way of computing
the correlation functions of twist operators, and use
a different way which we describe below. The usual computation for
$\langle\sigma\sigma\rangle $ (and the one adopted in \cite{dixon,frolov})
proceeds by
first finding
\be
f\equiv {\langle\partial X^i(z)\partial X^i(w)\sigma(z_1)\sigma(z_2)\rangle
\over
\langle\sigma(z_1)\sigma(z_2)\rangle }
\label{ten}
\ee
by looking at the singularities of $f$ as a function of $z,w$, and
constructing a function with these singularities. One then
takes the limit $z\rightarrow w$, subtracts the singularity and
constructs the stress tensor $T={1\over 2} \partial
X^i\partial X^i(w)$. Next, one uses the conformal Ward identity to
relate $\langle T\sigma\sigma\rangle $ to $\langle\partial\sigma
\sigma\rangle$,
thus obtaining an expression for $\partial_z
\log\langle\sigma(z)\sigma(0)\rangle $. Solving this equation gives the
functional
form of
the 2-point function, and the dimension of $\sigma$ can be read off
from the solution.

A similar analysis can be done for the 3-point function, but the
functional form of the 3-point function of primary fields is
determined by their dimensions, and would tell us nothing new. One
cannot find the fusion coefficients $C_{ijk}$ between
the twist operators from the 3-point analysis because the method does
not determine the overall normalization of the
correlator. Thus to find the $C_{ijk}$ one applies the method to the
4-point function $\langle\sigma\sigma\sigma\sigma\rangle $, finds
the functional form of this correlator, and then uses factorization
to extract the $C_{ijk}$.

To be able to use such a method one must have a simple stress tensor
which can be written as a product of fields, each of
which has a simple known behavior near the twist operators. One must
also use inspection to construct the correlators
like (\ref{ten}). Further, to find the $C_{ijk}$ we need to go up to
the 4-point function.

The method we use will apply  to  $S_N$ orbifolds, but not for
example to $Z_N$ orbifolds with $N>2$. On the other hand
we will not need that the stress tensor have a simple form (in fact
we will not use the stress tensor at all). Further, we can
compute the $C_{ijk}$ using only  the 2 and 3-point functions.  Thus
the method is suited to the computation of
correlation functions for CFTs arising from sigma models with target
space $M^N/S_N$, which arise in D-brae physics.  The method  also 
brings out the fact
that many quantities for symmetric orbifolds are `universal' in the
sense that they do not depend on the details of the the
manifold $M$.

\subsection{Closing the punctures}
\label{closing}

For  the discussion below the order and number of twist operators can
be arbitrary, but for explicitness we
assume that the covering space $\Sigma$ has the topology of a sphere, 
and we use the correlator
$\langle\sigma_2\sigma_2\rangle$ as an illustrative example. It will 
be evident that no new issues
arise for other correlators or when  $\Sigma$ has higher genus;  we 
will  mention  the
changes for higher genus where relevant.

As it stands the covering surface $\Sigma$ that we have constructed
has several `holes' in it. We will now
give a prescription for closing these holes, thus making a closed
surface which we also call $\Sigma$. The
prescription for closing the holes will amount  to defining
precisely the states  to be inserted at various
boundaries.  If the surface is closed then we can use the Liouville 
action to find the path
integral after change of metric; on an open surface the boundary 
states can change as well.

  The holes
are of the following kinds:

(i)\quad The holes in the finite $z$ plane at the insertion of the
twist operators. These holes are circles with
radius $\epsilon$ in the $z$ plane, and lift to holes in $\Sigma$
under the map $t(z)$.

(ii)\quad The holes in $\Sigma$ at finite values of $t$, arising from
the fact that we have cut the $z$ space at
$|z|=1/\delta$. These holes are located at points $t_0$ where the map
behaves as $z\sim {1\over t-t_0}$.

(iii)\quad The hole in $\Sigma$ at $t=\infty$, which also arises from
the fact that the $z$ space is cut at
$|z|=1/\delta$.  If there is no twist operator at $z=\infty$, then we 
have $z\sim t$ for large $t$,  and the
hole in $\Sigma$ is the image of $|z|=1/\delta$ under this map.

     We first  complete our
definition of the twist operators by defining  the state inserted at
the edge of the cut out
hole; this addresses holes of type (i) above.  We had said above that
the twist operator $\sigma_{12}$ imposes
the boundary condition (\ref{five}), but this does not
specify the operator completely. In fact there are an infinite number
of operators, with increasing dimensions, which all
create the same twist and in general create some further excitation
of the fields. We define $\sigma_{12}$  to be the operator  from this 
family with  lowest dimension. After
mapping the problem to the $t$ space, we need to ask what operator insertions
at the points $t=0$, $t=1$ give the slowest power law fall off for
the partition function when the distance $a$ between the
twist operators is increased. The answer is of course that we must
insert a multiple of the identity operator at the punctures in the $t$
space.

But we will also need to know the norm of this  state, and would thus
like to construct it through a path integral.
Thus let the covering surface be locally defined through
(\ref{sixp}). For $|t|>\epsilon^{1/2}$ the metric on the
covering surface is the one induced from the $z$ plane
\be
ds^2=dzd\bar z=dt d\bar t |{dz\over dt}|^2 =  4|t|^2dt d\bar t,
~~~~~|t|>\epsilon^{1/2}.
\ee
We `close the hole' in the $t$ space by choosing, for
$|t|<\epsilon^{1/2}$, the metric
\be
ds^2=4\epsilon dtd\tilde t,~~~|t|<\epsilon^{1/2}.
\label{sevenp}
\ee
Thus we have glued in a `flat patch'  in the $t$ space to close the
hole  created in the definition of the twist
operator. The metric is continuous across the boundary
$|t|=\epsilon^{1/2}$, but there is curvature
concentrated along this boundary. The path integral of $X$ over the
disc $|t|<\epsilon^{1/2}$ creates the required
state along the edge of the hole.  The map (\ref{sixp}) is only the
leading order approximation to the
actual map in general, but our prescription is to `close with a flat
patch' the hole in the $t$ space,
where the hole is the image on $\Sigma$ of a circular hole in the $z$
plane. As $\epsilon\rightarrow 0$, the
small departure of the map from the  form (\ref{sixp}) will cease to matter.

Note that we could have chosen a
different metric to replace the choice (\ref{sevenp}) inside the
hole, but this would just correspond to a
different overall normalization of the twist operator. (Thus it would
be  like taking a different choice of
$\epsilon$.)  Once we make the choice (\ref{sevenp}) then
we must use the same construction of the twist operator in all
correlators, and then the non--universal choices
in the definitions will cancel out.

The other holes, of types (ii) and (iii), arise from the hole at
infinity in the $z$ plane, and we proceed by first
replacing the $z$ plane by a closed surface.  We take another disc
with  radius $1/\delta$  (parameterized
by a coordinate $\tilde z$) and glue it to the boundary of the $z$
plane. Thus we get a sphere with metric given
by
\bea
d s^2&=&dzd\bar z, ~~~|z|<{1\over \delta},\nonumber \\
&=& d\tilde z d\bar{\tilde z},~~~|\tilde z|<{1\over \delta},\nonumber \\
\tilde z&=&{1\over \delta^2}{1\over z}.
\label{twp}
\eea

The path integral over the second disc defines a state at the
boundary of the first disc. This state is
proportional to the identity. But further,  our  explicit
construction gives the state a known norm, which is
something we needed to completely define the path integrals like
(\ref{four}) and (\ref{six}).

Since the $z$ space is closed at infinity, we find that the holes of
type (ii) and (iii) are now automatically closed
in $\Sigma$ --  since we make $\Sigma$ as a cover of the $z$ sphere
with metric on every patch induced
from the metric on this $z$ sphere.

The space $\Sigma$ is now a closed surface with a certain metric, and
the path integral giving $Z_{\epsilon,
\delta}[\sigma_2(z_1), \sigma_2(z_2)]$ in (\ref{six}) is to be
carried out on this closed surface.

\subsection{The method of calculation}

We had found above that if we have twist operators $\sigma_{12}$ at
$z=0$, $z=a$, then the partition function with the
twist operators inserted equals that for a single field $X$ on the
double cover $\Sigma$ of the $z$ plane given by the map
(\ref{eight}). $\Sigma$ is also a sphere with infinitesimal holes cut
out, but since only the identity operator is inserted at
these punctures, we  can close the holes and just get the partition
function of $X$ on
a closed surface  $\Sigma$.

At this point one might wonder that since this partition function is
some given number, how does it depend on the
parameter $a$, which gave the separation between the twist operators
in the $z$ plane? The point is that even though
$\Sigma$ is a sphere for all $a$, the {\it metric} on this sphere
depends on $a$ -- this is evident from (\ref{el}).

We {\it can} compute the partition function of $X$ on $\Sigma$ using
some fixed fiducial metric $\hat g$ on the $t$ space,
but we must then take into account the conformal anomaly, which says
that if $ds^2=e^\phi d{\hat s}^2$, then
the partition function $Z^{(s)}$ computed with the metric $ds^2$ is
related to the partition function $Z^{(\hat s)}$
computed with $d\hat s^2$ through
\be
Z^{(s)}=e^{S_L}Z^{(\hat s)},
\label{thir}
\ee
where
\be
S_L={c\over 96\pi}\int d^2t \sqrt{-g^{(\hat s)}}[ \partial_\mu\phi
\partial_\nu\phi g^{(\hat s)\mu\nu}+2R^{(\hat s)}\phi]
\label{fourt}
\ee
is the Liouville action \cite{friedan}.  Here $c$ is the central
charge of the CFT. Since we are considering  the theory of a
single free field $X$ on $\Sigma$, we have $c=1$.

Let us choose the fiducial metric $\hat g$ on $\Sigma$ to be (in the
case where $\Sigma$ is a sphere)
\bea
d\hat s^2&=&dtd\bar t, ~~~|t|<{1\over \delta'}\nonumber \\
&=& d\tilde t d\bar{\tilde t},~~~|\tilde t|<{1\over \delta' },\nonumber \\
\tilde t&=&{1\over \delta'^2}{1\over t}.
\label{tw}
\eea
We will let $\delta'\rightarrow 0$  at the end. Thus we have chosen 
the fiducial metric on
the
$t$ space to be the flat metric of a plane  up to a large radius
$1/\delta'$, after which we glue an identical disc at the
boundary to obtain the topology of a sphere, just as we did for the $z$ space.

    From (\ref{thir}), (\ref{fourt}) we see that if we increase $\phi$ by
a constant, then $Z$ changes by a known factor.
Using that fact that for a sphere $\int \sqrt{g}R=8\pi$, we find that
the partition function of $X$ on the sphere with metric (\ref{tw}) is
\be
Z_{\delta'}=Q~(\delta')^{-{c\over 3}}=Q~(\delta')^{-\frac{1}{3}}.
\label{mone}
\ee
Thus we will have $Z^{(\hat s)}=Z_{\delta'}$. Here $Q$ is a
constant that is regularization dependent and
cannot  determined by anything that we have
chosen so far. ($Q$ determines the size of the sphere  for which  the
partition function will attain the value unity;
since the CFT has no inbuilt scale we cannot find the value of this
size in any absolute way.)  $Q$ will
cancel out in all final calculations.

The partition function of one boson on the $z$ sphere with the metric
(\ref{twp})
is $Z_\delta$ (cf. eq. (\ref{four})), and we
have
\be
    Z_{\delta}=Q\delta^{-{c\over 3}}=Q\delta^{-{1\over 3}}.
\label{monep}
\ee

\subsection{Contributions to $S_L$}

The partition function with twist operators inserted  can be written as
\be
Z_{\epsilon, \delta}[\sigma_{n_1}(z_1),\dots, 
\sigma_{n_k}(z_k)]=e^{S_L}Z^{(\hat s)}.
\label{fift}
\ee
Thus the computation of the correlation function boils down to
computing $S_L$.  There are three types of contributions to $S_L$,
which we will analyze separately
\be
S_L=S_L^{(1)}+S_L^{(2)}+S_L^{(3)},
\ee
$S_L^{(1)}$ will give the essential numerical contributions to the
correlation functions (as well as regulation
dependent quantities), while $S_L^{(2)}$ and $S_L^{(3)}$ give only
regulation dependent quantities; regulation
parameters cancels out at the end.

(a)
\quad  We have cut out various discs from the
$z$ plane where the physical theory is defined: we have removed
infinity by  taking $|z|<1/\delta$ and have also cut
out circles of radius
$\epsilon$ around the twist operator insertions.  Let us call this
region of $z$ the `regular region'.  This
`regular'  region of the $z$ space has an image in the $t$ space,
which we call the `regular region' on $\Sigma$.
On $\Sigma$ we will find, apart from the obvious cuts around the
images of the twist operators and a cut
near
$|t|=\infty$, further possible  cuts around images of $z=\infty$
as discussed in subsection \ref{closing}. Let the contribution to
$S_L$ from this  `regular region'  of $\Sigma$ be called
$S_L^{(1)}$.

To evaluate (\ref{fift}) we need to choose a fiducial metric on the
$t$ space.
     Suppose that the map $z(t)$ has
the form
$z\sim bt$ as $t\rightarrow\infty$ .  (When there is no twist
operator at infinity the map can be taken to have
this form.) Let this fiducial metric
$d{\hat  s}^2$ be of the form (\ref{tw}) with
\be
{1\over \delta}<{b\over \delta'}.
\label{msix}
\ee
With this choice the boundary  $|z|=1/\delta$ gets mapped to a curve
inside the disc $|t|<1/\delta'$ (i.e. into the
`first half' of the $t$ sphere).

In this `regular region' of $\Sigma$,  the fiducial metric (\ref{tw})
is flat, and so there is
no contribution from the $R\phi$ term
in (\ref{fourt}). Thus we have
\be
S_L^{(1)}={1\over 96\pi}\int d^2t [\partial_\mu\phi \partial^\mu\phi],
\label{sixt}
\ee
where the integral extends over the region described above.
We rewrite (\ref{sixt}) as
\be
S_L=-{1\over 96\pi}\int d^2t [ \phi
\partial_\mu\partial^\mu\phi]+{1\over 96\pi}\int_{\partial}\phi
\partial_n\phi.
\label{sevt}\label{LiouvDef}
\ee
Here  $\partial$ is the boundary of the `regular region' of $\Sigma$,
and $\partial_n$ is the normal derivative at
the boundary. From (\ref{el}) we find that
\be
\phi=\log[{dz\over dt}] +\log[{d\bar z\over d\bar t}],
\label{eightt}
\ee
so that
\be
\partial_\mu\partial^\mu\phi=4\partial_t\partial_{\bar t}\phi =0,
\label{ninet}
\ee
and we get
\be
S_L={1\over 96\pi}\int_{\partial}\phi
\partial_n\phi.
\label{twenty}
\ee
The boundaries of the `regular region' are of two kinds: those arising from the
holes of size $|z-z_i|=\epsilon$ cut around the twist
operators, and those arising from the cutoff at infinity
($|z|=1/\delta$). Consider the boundary of the hole arising from
some twist operator $\sigma_n(z_i)$.  We regulated the twist operator 
by choosing this
hole to be a circle in the $z$ plane, so we start by
looking at a segment of the boundary using the coordinate $z$. We have
\be
\partial_n=-{1\over |z|}(z\partial_z+\bar z\partial_{\bar z}).
\label{tone}
\ee
Writing $z=|z|e^{i\theta}$, one finds
\be
\int ds =|z|\int d\theta ={|z|\over i}\int {dz\over z}={|z|\over
-i}\int {d\bar z\over \bar z}.
\label{ttwo}
\ee
Thus we get
\be
\int_{\partial} ds ~\phi \partial_n\phi=i\int dz ~\phi \partial_z\phi ~+~ c.c.
\label{tthree}
\ee
Since $z$ is holomorphically dependent on $t$, we can write
\be
dz~\partial_z\phi=dt~\partial_t\phi.
\label{tfour}
\ee
We can thus write for the contribution to $S_L$  from any hole
\be
{1\over 96\pi}\int_{\partial}ds~\phi
\partial_n\phi={1\over 96\pi}[i\int dt ~\phi \partial_t\phi ~+~ c.c.],
\label{tfive}
\ee
where $\phi$ is given through (\ref{eightt}). A similar analysis 
applies to all the other boundaries of the
`regular region' on $\Sigma$, and we compute (\ref{tfive}) for each 
such boundary.  Since the `holes' on
$\Sigma$ are infinitesimal size punctures, computing (\ref{tfive}) 
needs only the leading order
behavior of $\phi$ at the punctures.

(b)\quad We had cut out holes of radius $\epsilon$ in the $z$ plane
around the insertions of the twist
operators, and these gave corresponding holes in the `regular region'
of $\Sigma$.  We now compute the
contribution to $S_L$ from the part $H$ of $\Sigma$ that is used to
close such a hole.    Since we had
closed these holes with the flat metric (\ref{sevenp}), and since the
fiducial metric we use on
$\Sigma$ is also flat  in $H$ ($d{\hat s}^2=dtd\bar t$),  we get
$\phi={\rm constant}$, and so there is no
contribution from the kinetic term in (\ref{fourt}). Note that at the
boundary of $H$ we have $\partial_t\phi$
nonzero but bounded,  then since the area of this boundary is zero (the
boundary is one-dimensional) we get
no contribution to the kinetic term from the boundary either.  The
curvature term in (\ref{fourt}) is zero,
since the curvature of the fiducial metric is zero throughout the
region where the twist operators are
inserted.  Thus we get no contribution to $S_L$ from these regions
$H$ of $\Sigma$.

(c)\quad Now consider the contributions from the points that have
finite $t$, but $z\rightarrow\infty$. The
`regular region' on $\Sigma$ had excluded the image of
$|z|>1/\delta$. This image will have  a small disc
$D$ around some finite $t_0$, if we have
\be
z\approx {\alpha\over t-t_0}+\beta+\dots
\label{mseven}
\ee
The fiducial metric we are using on $\Sigma$ is flat here, so there
is no contribution from the curvature term
in (\ref{fourt}).  The region inside the disc $D$ has a metric
induced from the `second half' of the $z$ sphere
(i.e. the part parameterized by $\tilde z$ in (\ref{twp})) so that the
metric is $ds^2=d\tilde zd{\bar{\tilde z}}$.
Thus
\bea
\tilde z&=&{1\over \delta^2}{1\over z}\approx {1\over
\delta^2\alpha}(t-t_0)-{\beta\over
\delta^2\alpha^2}(t-t_0)^2,\nonumber\\
\phi&=&\log{d\tilde z\over dt}+c.c.\approx \log{1\over
\delta^2\alpha}-{2\beta\over
\alpha}(t-t_0)+c.c.,\nonumber\\
\partial_t\phi&\approx& -{2\beta\over \alpha}.
\label{meight}
\eea
The area of the disc $D$ in the fiducial metric is $\pi
|t-t_0|^2\approx \pi(|\alpha|\delta)^2$.  As
$\delta\rightarrow 0$, we find that $\int d^2t ~
\partial_t\phi\partial_{\bar t}\phi\rightarrow 0$. Thus we get no
contribution to $S_L$ from these images of the cut at infinity.

(d)\quad Now we look at the region of $\Sigma$ near $t=\infty$. Let
$z\approx bt$ for large $t$.
Let the image of $|z|=1/\delta$ be the contour $C$ on $\Sigma$. By
the choice (\ref{msix}) and the fact that
$C$  satisfies  $|t|\approx {1\over b\delta}$,  we find that $C$ is
inside the curve
$|t|=1/\delta'$. Let the contribution to $S_L$ from the region
between $C$ and $|t|=1/\delta'$
be called $S_L^{(2)}$.

Since the fiducial metric (\ref{tw}) is flat in this region, there is
no contribution from the curvature term in
(\ref{fourt}).  For the kinetic term we have
\bea
\tilde z&=&{1\over \delta^2 z}\approx {1\over \delta^2 bt},
~~~{d\tilde z\over dt}\approx -{1\over
\delta^2 b}{1\over t^2},\nonumber\\
\partial_t\phi&\approx&-{2\over t},\qquad
\int d^2 t \partial\phi\partial\phi\approx
32\pi\log{\delta'\over |b|\delta}.
\eea
and we get
\be
S_L^{(2)}=-{1\over 3}\log{\delta'\over |b|\delta}.
\label{mten}
\ee

(e)\quad Moving further outwards in the $t$ plane, we find a `ring of 
curvature' at
$|t|=1/\delta'$  (cf. eq.
(\ref{tw})).  At this ring we have
\bea
\phi\approx \log{(\delta')^2\over
\delta^2 b}+c.c.,\qquad
\int d^2t R\phi=8\pi\phi,
\eea
which gives a contribution to $S_L$ equal to
\be
S_L^{(3)}={1\over 3} \log{(\delta')^2\over
\delta^2 |b|}.
\label{mel}
\ee
The kinetic term in $\phi$ has no contribution at this ring. Further,
the region $|t|>1/\delta$ gives
no contribution to $S_L$, since the curvature of the fiducial metric
is zero, and the map gives $\phi={\rm constant}$.

\subsection{The correlator in terms of the Liouville field}

Let us collect all the above contributions together.
Note that
\be\label{LiouvIfty}
S_L^{(2)}+S_L^{(3)}=\frac{1}{3}\log\frac{\delta'}{\delta}.
\ee
so that the variable $b$ drops out of this combination.

Now let us go back to the expression (\ref{six}) that we want to evaluate :
\be\label{6.23.twoPt}
\langle\sigma_n^\epsilon(0)\sigma_n^\epsilon(a)\rangle_\delta=
{Z_{\epsilon,
\delta}[\sigma_n(z_1), \sigma_n(z_2)]\over (Z_\delta)^N}=
e^{S_L}{Z^{(\hat s)}\over (Z_\delta)^n},
\ee
here we used equation (\ref{thir}). Taking into account the relations
(\ref{mone}) and (\ref{monep}) we finally get:
\be
\langle\sigma_n^\epsilon(0)\sigma_n^\epsilon(a)\rangle_\delta=
e^{S_L}\left(\frac{\delta^n}{\delta'}\right)^{1/3}Q^{1-n}.
\ee
Substituting the expression for the Liouville action, $S_L$, we conclude that
\be\label{ActionS4}
\langle\sigma_n^\epsilon(0)\sigma_n^\epsilon(a)\rangle_\delta=
e^{S_L^{(1)}}e^{S_L^{(2)}+S_L^{(3)}}
\left(\frac{\delta^n}{\delta'}\right)^{1/3}Q^{1-n}=
e^{S_L^{(1)}}\delta^{\frac{n-1}{3}}Q^{1-n}.
\ee
Thus we observe a cancellation of $\delta'$, which served only to
choose a  fiducial metric on $\Sigma$ and thus
should not appear in any final result. The only quantity that needs
computation is $S_L^{(1)}$ using (\ref{tfive}).

Let us mention that formula (\ref{6.23.twoPt}) has a simple extension to the
case of a general correlation function:
\be\label{TheCorrelator}
\langle\sigma_{n_1}^\epsilon(z_1)\dots
\sigma_{n_k}^\epsilon(z_k)\rangle_\delta=
e^{S_L}{Z^{(\hat s)}\over (Z_\delta)^s},
\ee
where $Z^{(\hat s)}$ is a partition function of the covering Riemann
surface $\Sigma$ with the fiducial metric $d\hat s^2$ ($\Sigma$ may
have any genus),
and $s$ is a number of fields involved in nontrivial permutation ($s=n$ in the
case of the two point function (\ref{6.23.twoPt})). The
partition function $Z^{(\hat s)}$ may depend on the moduli of the surface
$\Sigma$
and its size (there are no moduli in the case of the sphere and the size is
parameterized by $\delta'$).

\section{The 2-point function}
\label{Section2pt}
\renewcommand{\theequation}{3.\arabic{equation}}
\setcounter{equation}{0}
\subsection{The calculation}

Let us apply the above scheme to evaluate the
2-point function of twist operators.  If one of the twist operators
corresponds to the
permutation
\be
(1\dots n),
\ee
then the other one should correspond to the permutation
\be
(n\dots 1),
\ee
since otherwise the correlation function vanishes. Thus we can write
\be
\langle \sigma_n (0)\sigma_n (a)\rangle
\label{tsix}
\ee
instead of $
\langle \sigma_{(1\dots n)} (0)\sigma_{(n\dots 1)} (a)\rangle$
without causing confusion.

The generalization of the map (\ref{eight}) to the case of $\sigma_n$ is
\be
z=a{ t^n\over t^n-(t-1)^n}.
\ee
For this map we have
\bea
\phi&=&\log|{dz\over dt}|^2=\log[an{t^{n-1}(t-1)^{n-1}\over
(t^n-(t-1)^n)^2}]+c.c.,\\
{d\phi\over dt}&=&
-{(2t+n-1)(t-1)^n-(2t-n-1)t^n\over t(t-1)((t-1)^n-t^n)}\nonumber
\eea
This map has the branch points located at
\be
t=0 \rightarrow z=0\qquad \mbox{and}\qquad
t=1\rightarrow z=a.
\ee
There are $n$ images of the point $z=\infty$ in $t$ plane:
\be
t_k={1\over 1-\alpha_k},\qquad \alpha_k=e^{2\pi i k\over n},
~~~~~~k=0,1,\dots , n-1.
\ee
We note that $\alpha_0=1$ gives $t=\infty$.

Let us compute the contribution (\ref{tfive}) for the point $z=0$.
Near this point we have:
\bea
z\approx (-1)^{n+1}at^n, \qquad |t|\approx{|z|^{1/n}\over a^{1/n}},\\
\phi\approx \log[ant^{n-1}],\qquad
\partial_t\phi\approx {n-1\over t}.\nonumber
\eea
Then we get the contribution to the Liouville action (\ref{LiouvDef}):
\be
S_L(t=0)=-\frac{n-1}{12}\left[{1\over n}\log |a| +
\log(n \epsilon^{n-1\over n})\right].
\ee
By a reflection symmetry $t \rightarrow 1-t$, $z\rightarrow a-z$, we
get the same contribution from the other branch point:
\be
S_L(t=1)=-\frac{n-1}{12}\left[{1\over n}\log |a| +
\log(n \epsilon^{n-1\over n})\right].
\ee

Now we look at the images of infinity.
First we note that the integral over the boundary located near $t=\infty$
will give zero, since $d\phi/dt$ goes like $1/t^2$,  the length of the
circle goes like $t$ and the value of $\phi$ is at best logarithmic in $t$.
But we do get a contribution from  the images of $z=\infty$  located
at finite points in $t$ plane.
Note that
\be
\mbox{if}\quad t={1\over 1-\alpha_k}+x,\quad \mbox{then}\quad
(t-1)^n-t^n\approx {xn\over \alpha_k(1-\alpha_k)^{n-2}}.
\ee
This leads to
\bea
z\approx -\frac{a\alpha_k}{n(1-\alpha_k)^2}\ \frac{1}{x},\qquad
x\approx -\frac{a\alpha_k}{n(1-\alpha_k)^2}\ \frac{1}{z},\\
\phi=\log\left[a^{-1}n(1-\alpha_k)^2 \alpha_k^{n-1}z^2\right],\qquad
\partial_t\phi\approx-\frac{2}{t-(1-\alpha_k)^{-1}} .
\eea
The point $t=t_k$ we are considering gives following contribution to the
Liouville action:
\be
S_L(t=t_k)=\frac{1}{6}\log\left[a^{-1}n(1-\alpha_k)^2 \alpha_k^{n-1}
{\delta}^{-2}\right].
\ee
Thus the total contribution from the images of infinity is:
\be
S_L(z=\infty)=\frac{1}{6}\sum_{k=1}^{n-1}
\log\left[\frac{n(1-\alpha_k)^2 \alpha_k^{n-1}}{a{\delta}^{-2}}
\right]=
-\frac{n-1}{6}\log[{|a|{\delta}^2}]+\frac{n+1}{6}\log[n].
\ee
We have used the following properties of $\alpha_k$:
\be
\prod_{k=1}^{n-1}\alpha_k=1;\qquad
\prod_{k=1}^{n-1}(q-\alpha_k)=\frac{q^n-1}{q-1}\rightarrow n,\quad  \mbox{if}
\quad q\rightarrow 1,
\ee
which follow from the fact that $\{\alpha_k\}$ is the set of different
solutions of the equation $\alpha^n-1=0$ and $\alpha_0=1$.

Adding all the contributions together, we get an expression for the
interesting part of the Liouville
action:
\be
S^{(1)}_L= -\frac{1}{6}\left[(n-{1\over n})\log|a|+{(n-1)^2\over
n}\log\epsilon+2(n-1)\log{\delta} -2\log n\right].
\ee
This leads to the final expression for the correlation function
(see (\ref{ActionS4})):
\be\label{2ptfunction}
\langle \sigma^\epsilon_n (0)\sigma^\epsilon_n (a)\rangle_\delta=
e^{S^{(1)}_L}\delta^{\frac{n-1}{3}}Q^{1-n}=
a^{-{1\over 6}(n-{1\over n})}~C_n\epsilon^{A_n}Q^{B_n},
\ee
\be\label{2ptfuncoef}
A_n=-{(n-1)^2\over 6n},\quad B_n=1-n,\quad
C_n=n^{1/3}.
\ee
Thus we read off the dimension $\Delta_n$ of $\sigma_n$
\be
\Delta_n={1\over 24}(n-{1\over n})
\label{tseven}
\ee
The other constants in (\ref{2ptfunction}) are to be absorbed into
the normalization of $\sigma_n$. We will discuss this
renormalization after computing the 3-point functions.

\subsection{`Universality' of the 2-point function}

The theory we  have considered above is that of the orbifold
$M^N/S_N$ where the manifold $M$ is just $R$, the real line.
If $M$ was $R^d$ instead,  we could treat the $d$ different species
of fields independently, and obtain
\be
\Delta_n={c\over 24}(n-{1\over n})
\label{teight}
\ee
where $c=d$ is the central charge of the CFT for one copy of $M=R^d$.
But we see that we would obtain the result
(\ref{teight}) for the symmetric orbifold with {\it any} choice of
$M$; we just use the value of $c$ for the CFT on $M$.
Around the insertion of the twist operator we permute the copies of
$M$, but the definition of the twist operator does not
involve directly the structure of $M$ itself.  The Liouville action
(\ref{fourt}) determines the correlation function using only
the value $c$ of the CFT. Thus we recover the result (\ref{teight})
for any $M$.

This `universality' of $\Delta_n$ is well known, and the value of
$\Delta_n$ can be deduced from the following  standard argument.
   Consider the CFT on a cylinder parameterized by
$w=x+iy, ~0<y<2\pi$. At $x\rightarrow -\infty$ let the state
be the vacuum of the orbifold CFT $M^N/S_N$. Since there is no twist,
   each copy of $M$ gives its own contribution
to the vacuum energy, which thus equals $-{c\over 24}N$.  Now insert
the twist operator $\sigma_n$ at $w=0$, and look at
the state for $x\rightarrow\infty$. The copies of $M$ not involved in
the twist contribute $-{c\over 24}$ each as before, but
those that are twisted by $\sigma_n$ turn into  effectively one copy
of $M$ defined on a circle of length $2\pi n$. Thus the
latter set contribute $-{c\over 24 n}$ to the vacuum energy. The
change in the energy between $x = \infty$ and
$x=-\infty$ gives the dimension of $\sigma_n$ (since the state at
$x\rightarrow -\infty$ is the vacuum)
\be
-{c\over 24 n} -[-{cn\over 24}]={c\over 24}(n-{1\over n})=\Delta_n
\ee
Thus while our calculation of the 2-point function has not taught us
anything new,  we have obtained a scheme that will
yield the higher point functions for symmetric orbifolds using an
extension of the same universal features that gave the
value of $\Delta_n$ in the above argument.

\section{The map for the 3-point function}
\renewcommand{\theequation}{4.\arabic{equation}}
\setcounter{equation}{0}
\subsection{Genus of the covering surface}

Let us first discuss the nature of the covering surface $\Sigma$ for
the case where we have an arbitrary number of twist
operators in the correlation function $\langle\sigma_{n_1}\sigma_{n_2}\dots
\sigma_{n_k}\rangle $.  The CFT is still defined on the
plane $z$, which we will for the moment regard as a sphere by
including the point at infinity.

At the insertion of the operator $\sigma_{n_j}(z_j)$ the covering
surface $\Sigma$ has a branch point of order $n_j$, which
means that $n_j$ sheets of $\Sigma$ meet at $z_j$. One says that the
ramification order at $z_j$ is $r_j=n_j-1$. Suppose
further that over a generic point $z$ here are $s$ sheets of the
covering surface $\Sigma$. Then the genus $g$ of $\Sigma$
is given by the Riemann--Hurwitz formula:
\be
g~=~\frac{1}{2}\sum_j r_j - s + 1
\label{tnine}
\ee

Let us now consider the 3-point function.  We require each twist operator to
correspond to a single cycle of the permutation group,
and regard the product of two cycles to represent the
product of two different twist operators.  Let the cycles
have lengths $n,m,q$ respectively.  It is easy to see that we can
obtain covering surfaces $\Sigma$ of various genera.
For example, if we have
\be
\sigma_{12}~~\sigma_{13} ~ ~\sigma_{123}
\ee
as the three permutations, then we have $r_1=1, r_2=1, r_3=2, s=3$,
and we get $g=0$. On the other hand with
\be
\sigma_{123}~~\sigma_{123}~~\sigma_{123}
\ee
we get $r_1=r_2=r_3=2, s=3$ and we get $g=1$.  (This genus 1 surface
is a singular limit of the torus, however.)

Let us concentrate on the case where we get $g=0$.
Without loss of generality we can take the first permutation
$\sigma_n$ to be the cycle
\be
(1,~2,~\dots ~k,~k+1,~\dots ~n)
\label{thone}
\ee
The second permutation is restricted by the requirement that when
composed with (\ref{thone}) it yields a single cycle
(which would be the conjugate permutation of the third twist
operator).  In addition we must have a sufficiently small
number of indices in the result of the first two permutations so that
we do get $g=0$. A little inspection shows that
$\sigma_m$ must have the form
\be
(k,~k-1,~\dots~1,~n+1,~n+2,~\dots~n+m-k)
\label{thtwo}
\ee
Thus the elements $1,~2,~\dots~k$ of the first permutation occur in the
second permutation in the reverse order, and
then we have a new set of elements $n+1,~\dots ~n+m-k$. These two
permutations compose
to give the cycle
$\sigma_m\sigma_n$ equal to
\be
(~k+1,~k+2,~\dots~ n,~1,~n+1,~n+2,~\dots ~ n+m-k)
\label{ththree}
\ee
Thus $\sigma_q$ must be the inverse of the cycle (\ref{ththree}), and we have
\be
q=n+m-2k+1
\label{thfour}
\ee
Note that the number of `overlaps' (i.e., common indices) between
$\sigma_n$ and $\sigma_m$ is $k$. Note that we must
have
$k\ge 1$ in order that the product $\sigma_m\sigma_n$ be a single
cycle rather than just a product of two cycles. Also note that if we 
have $q=n+m-1$,  then
since $s\ge q$, (\ref{tnine}) gives that $\Sigma$ must have genus 
zero (this will be a `single overlap'
correlator).

Let $\Sigma$ be the covering surface that corresponds to the insertions \\
$\langle\sigma_n(z_1)\sigma_m(z_2)\sigma_q(z_3)\rangle$. Then  the
number of sheets of $\Sigma$ over a generic point $z$
is just the total number of indices used in the permutations
\be\label{NumberOfSheets}
s=n+m-k={1\over 2}(n+m+q-1)
\label{thfive}
\ee
Thus the genus of $\Sigma$ is
\be\label{GenusG=0}
{n-1\over 2}+{m-1\over 2}+{q-1\over 2}-s+1=0
\label{thsix}
\ee

\subsection{The map for the case $g=0$}

We are looking for a covering surface of the sphere that is ramified
at three points on the sphere, with a finite order of
ramification at each point.  We look for the map from $z$ to
$\Sigma$  as a ratio of two polynomials
\be
z={f_1(t)\over f_2(t)};
\label{thseven}
\ee
the existence of such a  map will be evident from its explicit construction.
By using the $SL(2,C)$ symmetry group of the $z$ sphere, we will
place the twist operators $\sigma_n, \sigma_m, \sigma_q$
at
$z=0, z=a, z=\infty$ respectively.  We can assume without loss of
generality that
\be\label{orderchoise}
n\le q, \qquad m\le q.
\ee
Note that we had placed a cutoff in the $z$
plane to remove the region at infinity, and  will not be immediately
clear how to normalize a twist that occurs around  the
circle at infinity.  We will discuss this issue of normalization later.

By making an $SL(2,C)$ transformation $t'=\frac{at+b}{ct+d}$ of the
surface $\Sigma$, which we assume is parameterized by
the coordinate $t$, we can take
\be
z(t=0)=0, \qquad z(t=\infty)=\infty, \qquad z(t=1)=a,
\label{theight}
\ee
Note that this $SL(2,C)$ transformation maintains the form
(\ref{thseven}) of $z$ to be a ratio of two polynomials, and we
will use the symbols $f_1, f_2$ to denote the polynomials after the
choice  (\ref{theight}) has been made.

Since we need  $s$ values of $t$ for a generic value of $z$, with
$s$ given by (\ref{thfive}), the relation (\ref{thseven})
should give a polynomial equation of order $s$ for $t$.
Thus the  degrees  $d_1, d_2$ of the polynomials $f_1, f_2$ should satisfy:
$$
\max(d_1,d_2)=s=\frac{1}{2}(n+m+q-1).
$$
Since we have chosen $t=\infty$ for $z=\infty$, we get $d_1>d_2$,  and we
have
\be\label{expr_d1}
d_1=\frac{1}{2}(n+m+q-1).
\ee
The requirement of the proper behavior at infinity ($z\sim t^q$) then gives:
\be\label{expr_d2}
d_2=d_1-q=\frac{1}{2}(n+m-q-1).
\ee
Finally, the the number of indices common between the permutations
$\sigma_n(0)$ and
$\sigma_m(a)$  (the overlap) is
\be
\frac{1}{2}(n+m-(q-1))=d_2+1.
\ee

Let us now look at the structure required of the map (\ref{thseven}).
For $z\rightarrow 0$ we need
\be
z=t^n(C_0+O(t))
\ee
      Similarly for $t\rightarrow 1$ we need
\be
z=a+(t-1)^m(C_1+O(t-1))
\ee
Then we find
\be
f_2^2\frac{dz}{dt}=f'_1f_2-f'_2f_1=C t^{n-1}(1-t)^{m-1}.
\label{thnine}
\ee
($C$ is a constant). The last step follows on noting that the
expression $f'_1f_2-f'_2f_1$ is a polynomial of degree
      $d_1+d_2-1=n+m-2$, and the behavior of $z$  near $z=0, z=a$ already
provides all the possible zeros of this polynomial
$f_2^2\frac{dz}{dt}$. The expression in (\ref{thnine}) is just the
Wronskian of $f_1, f_2$, and our knowledge of this
Wronskian given an easy way to find these polynomials.  We seek a
second order linear differential equation whose
solutions are the linear span  $f=\alpha f_1+\beta  f_2$. Such an
equation is found by observing that
\be
\left|\begin{array}{ccc}
f&f'&f''\\
f_1&f_1'&f_1''\\
f_2&f_2'&f_2''
\end{array}\right|=0
\ee
so that we get the equation
\be
Wf''-W'f'+c(t) f=0
\label{fone}
\ee
where
\be
W=f_2f_1'-f_1f_2', ~~~c(t)=f_2'f_1''-f_1'f_2''
\label{ftwo}
\ee
Here $W$ is given by (\ref{thnine}). The coefficient $-W'$ of $f'$ is
\be
-W'=-Ct^{n-2}(t-1)^{m-2}[(n-1)-(n+m-2)t]
\label{fthree}
\ee
The coefficient $c(t)$ must be a polynomial of degree $n+m-4$ but in
fact we can argue further  that it must have the
form
\be
\gamma ~t^{n-2}(1-t)^{m-2}, ~~~\gamma={\rm constant}
\label{ffour}
\ee
To see this look at the equation near $t=0$. Let $c(t)\sim \alpha t^k$ with
$k<n-2$.  Then the equation reads
\be
t^{n-1-k} f''-(n-1)t^{n-2-k}f'+{\alpha\over C}f=0
\label{ffive}
\ee
Note that the two polynomials $f_1, f_2$ which solve the equation
must not have a common root $t=0$, since we assume
that (\ref{thseven}) is already expressed in reduced form. Thus at
least one of the solutions must go like $f\sim constant$ at
$t=0$, which is in contradiction with (\ref{ffive}) since the first
two terms on the LHS vanish while the last does not ($a\ne
0$ by definition).  Thus $c(t)$ has a zero of order at least $n-2$ at
$t=0$, and by a similar argument, a zero of order at least
$m-2$ at $t=1$. Thus the result (\ref{ffour}) follows.

Dividing through by $Ct^{n-2} (1-t)^{m-2}$ we can write the equation
(\ref{fone}) as
\be
t(1-t)f''+[-(n-1)+(n+m-2)t]f'+\tilde \gamma=0
\ee
Let us now look at $t\rightarrow\infty$, and let the solutions to the
above equation go like $t^p$. Then we get
\be
-p(p-1)+p(m+n-2)+\tilde\gamma=0.
\ee
which has the solutions
\be
p_\pm=\frac{1}{2}\left(m+n-1\pm\sqrt{(m+n-1)^2+4\tilde\gamma}\right)
\ee
But since we have a twist operator of order $q$ at infinity, we must have
\be
p_+-p_-=q,
\ee
This gives
\be
\tilde\gamma=\frac{1}{4}(q-m-n+1)(q+m+n-1)=-d_1d_2.
\ee

Thus we have found the equation which is satisfied by both $f_1$ and $f_2$:
\be\label{hypereqn}
t(1-t)y''+(-n+1-(-d_1-d_2+1)t)y'-d_1d_2y=0,
\ee
which is  the hypergeometric equation. Its general solution is given by
\be\label{hypersolution}
y=AF(-d_1,-d_2;-n+1;t)+Bt^nF(-d_1+n,-d_2+n;n+1;t).
\ee
     The map we are looking for can be written as
\be\label{maphyper}
z=a\frac{d_2!d_1!}{n!(d_1-n)!}\frac{\Gamma(1-n)}{\Gamma(1-n+d_2)}t^n
\frac{F(-d_1+n,-d_2+n;n+1;t)}{F(-d_1,-d_2;-n+1;t)},
\ee
where we have chosen the normalizations of $f_1, f_2$ such that the
$t=1$  maps to $z=a$.  In our case $d_1$, $d_2$ and $n$ are integers.
Some of the individual terms in
the above expression are undefined for integer $d_1$, $d_2$, $n$ and
a limit should be taken from non--integer
values of $n$ (while keeping $d_1, d_2$ fixed at their integer
values). We can write the result in a
well defined way by using  Jacobi polynomials, which are a set of
orthogonal polynomials defined through the
hypergeometric function
\bea
P_n^{(\alpha,\beta)}(x)&\equiv&\left(\begin{array}{c}n+\alpha\\n\\\end
{array}\right)
F(-n,n+\alpha+\beta+1;\alpha+1;{1-x\over
2})\nonumber \\
&=&{1\over n!}\sum_{\nu=0}^n
\left(\begin{array}{c}n\\
\nu\\\end{array}\right)(n+\alpha+\beta+1)\dots(n+\alpha+\beta+\nu)
\nonumber \\
&&\cdot(\alpha+\nu+1)\dots(\alpha+n)({x-1\over 2})^\nu
\label{mthree}
\eea
Then (\ref{maphyper}) becomes
\be\label{mapjac}
z=at^nP_{d_1-n}^{(n,-d_1-d_2+n-1)}(1-2t)
\left[P_{d_2}^{(-n,-d_1-d_2+n-1)}(1-2t)\right]^{-1}.
\label{sthree}
\ee

We will have occasion to use the Wronskian of the polynomials later,
and we define $\tilde W$ to be normalized as
follows
\bea\label{WronskJacobi}
{\tilde W}(t)&=&\frac{d}{dt}\left[t^nP_{d_1-n}^{(n,-d_1-d_2+n-1)}(1-2t)\right]
P_{d_2}^{(-n,-d_1-d_2+n-1)}(1-2t)\nonumber\\
&-&t^nP_{d_1-n}^{(n,-d_1-d_2+n-1)}(1-2t)\frac{d}{dt}
P_{d_2}^{(-n,-d_1-d_2+n-1)}(1-2t)\nonumber\\
&=&\frac{nd_1!}{n!d_2!(d_1-n)!}
\frac{\Gamma(d_2-n+1)}{\Gamma(1-n)}t^{n-1}(1-t)^{d_1+d_2-n},
\eea
We will also have occasion to use the relation  (\ref{maphyper}) 
containing hypergeometric functions,
and we define
\bea\label{WronskSolut}
W(t)&=&\left[t^nF(-d_1+n,-d_2+n;n+1;t)\right]'F(-d_1,-d_2;-n+1;t)\nonumber\\
&-&t^nF(-d_1+n,-d_2+n;n+1;t)F'(-d_1,-d_2;-n+1;t)\nonumber\\
&=&nt^{n-1}(1-t)^{d_1+d_2-n}.
\eea
We will calculate the three point
function using the map (\ref{maphyper}), (\ref{mapjac}) in the next section.

\section{The Liouville action for the 3-point function}
\renewcommand{\theequation}{5.\arabic{equation}}
\setcounter{equation}{0}
Let us evaluate the three point function
\be
\langle \sigma_n(0) \sigma_m(a)\sigma_q(\infty)\rangle
\ee
using the map (\ref{maphyper}), (\ref{mapjac}).

Recall that we  cut circles of radius $\epsilon$ in the $z$ plane
around the twist operators at $z=0$ and $z=a$ to regularize
these twist operators.  But unlike the case of the
2-point function discussed in section \ref{Section2pt}, now we have the  twist
operator $\sigma_q$  inserted at infinity. This means
that the fields $X^I$ have boundary conditions around  $z=\infty$ 
such that $q$ of the $X^I$ form a cycle
under rotation around the circle $X^{i_1}\rightarrow
X^{i_2}\rightarrow\dots X^{i_q}\rightarrow X^{i_1}$, while the
remaining fields $X^I$ are single valued around this circle. Note
that if the covering surface $\Sigma$ has $s$ sheets over a
generic $z$ then there will be $s-q$ such single valued fields $X^I$.

The covering  surface $\Sigma$  will have punctures at $t=0$ and
$t=1$ corresponding to $z=0$ and $z=a$ respectively. In
addition it will have  punctures corresponding to the `puncture at
infinity'  in the $z$ plane.
These latter punctures are of two kinds. The first kind of puncture
in the $t$ plane will correspond to the place where $q$
sheets meet in the $z$ plane - i.e., the lift of the point where the
twist operator was inserted. But we will also have $s-q$
other punctures in the $t$ plane that correspond to the cut at
$|z|=1/\delta$ for the $X^I$ that are single valued around
$z=\infty$. We will choose (when defining the `regular region') a 
cutoff at value $|z|=1/\tilde\delta$ for
the first kind of puncture  (i.e. the puncture arising from fields 
$X^I$ that are twisted at $z=\infty$) and a
value
$|z|=1/\delta$   for the second kind of puncture (i.e. punctures for fields
$X^I$ which are not twisted at infinity).  We will see that both
$\delta$ and
$\tilde\delta$ cancel  from all final results.

\subsection{The contribution from  $z=0, t=0$}

Let us first consider the point  $z=0$ which gives $t=0$.
Near this point the map (\ref{maphyper}) gives:
\be
z\approx a\frac{d_2!d_1!}{n!(d_1-n)!}\frac{\Gamma(1-n)}{\Gamma(1-n+d_2)}t^n,
\qquad
t\approx\left(\frac{z n!(d_1-n)!
\Gamma(1-n+d_2)}{ad_2!d_1!\Gamma(1-n)}\right)^{1/n}.
\ee
Note that by using the relation
\be
\Gamma(x)\Gamma(1-x)={\pi\over \sin(\pi x)}
\label{sone}
\ee
we can write
\be
{\Gamma(1-n)\over \Gamma(1-n+d_2)}={\Gamma(n-d_2)\over
\Gamma(n)}{\sin(\pi(n-d_2))\over \sin(\pi
n)}={(n-1)!\over (n-d_2-1)!}(-1)^{d_2}\label{stwo},
\ee
so that the $\Gamma$ functions in the above expressions are in reality
well defined.

The Liouville field and its derivative are given by:
\be
\phi\approx\log\left(\frac{nad_2!d_1!}{n!(d_1-n)!}
\frac{(n-d_2-1)!}{(n-1)!}t^{n-1}\right)+c.c., \qquad
\partial_t\phi\approx\frac{n-1}{t},
\ee
where we have dropped the factor $(-1)^{d_2}$ in (\ref{stwo}) since
$\phi$ is the real part of the logarithm.
Substituting these values into the expression for the Liouville action:
\be\label{s3ptLiouv}
S_L=\frac{i}{96\pi}\int dt\phi\partial_t\phi,
\ee
we get a contribution from the point $t=0$:
\be
S_L(t=0)=-\frac{n-1}{12}\log\left(n\epsilon^{\frac{n-1}{n}}\right)-
\frac{n-1}{12n}\log\left(a\frac{d_2!d_1!}{n!(d_1-n)!}\frac{(n-d_2-1)!}{
(n-1)!}\right),
\ee
where we note that the integration in (\ref{s3ptLiouv}) is performed
along the circle
\be
|t|=\left(\frac{\epsilon n!(d_1-n)!
(n-1)!}{ad_2!d_1!(n-d_2-1)!}\right)^{1/n}.
\ee

A simplification analogous to (\ref{stwo}) will occur in many
relations below, but for simplicity we leave the $\Gamma$
functions in the form  where they have negative arguments; we
replace them with factorials of positive numbers
only in the final expressions.

\subsection{The contribution from  $z=a, t=1$}

      Let us look at the point $t=1$. Using the expression for Wronskian
(\ref{WronskSolut}), we find the derivative of the map (\ref{maphyper}):
\be
\frac{dz}{dt}=a\frac{d_2!d_1!}{n!(d_1-n)!}\frac{\Gamma(1-n)}{\Gamma(1-n+d_2)}
\frac{nt^{n-1}(1-t)^{d_1+d_2-n}}{\left[F(-d_1,-d_2;-n+1;t)\right]^2},
\ee
which can be combined with known property of hypergeometric function:
\be
F(a,b;c;1)=\frac{\Gamma(c)\Gamma(c-a-b)}{\Gamma(c-a)\Gamma(c-b)}
\ee
to give the result:
\bea
z&\approx& a-\beta a(1-t)^{d_1+d_2-n+1},\\
\label{betafactor}
\beta&=&\frac{n}{d_1+d_2-n+1}\ \frac{d_1!d_2!(d_1-n)!}{n!
\left[(d_1+d_2-n)!\right]^2}\
\frac{\Gamma(1-n+d_2)}{\Gamma(1-n)}.
\eea
Our usual analysis gives:
\bea
z&\approx& a-\beta a(1-t)^{d_1+d_2-n+1},\qquad
1-t\approx\left(-\frac{z-a}{a\beta}\right)^{\frac{1}{d_1+d_2-n+1}},\nonumber\\
\phi&\approx&\log\left(a\beta(d_1+d_2-n+1)(1-t)^{d_1+d_2-n}\right)+c.c.,\quad
\partial_t\phi\approx -\frac{d_1+d_2-n}{1-t},\nonumber
\eea
\bea
S_L(t=1)&=&-\frac{d_1+d_2-n}{12}\log(d_1+d_2-n+1)\\
&-&\frac{(d_1+d_2-n)^2}{12(d_1+d_2-n+1)}\log\epsilon-
\frac{(d_1+d_2-n)}{12(d_1+d_2-n+1)}\log(a|\beta|).\nonumber
\eea

Note that $d_1+d_2-n+1=m$, so that we can rewrite the contribution
from $t=1$ in a way which makes it look more
symmetrical with the contribution from $t=0$. But we will defer all
such simplifications to the final expressions for the
fusion coefficients.

\subsection{The contribution from $z=\infty$}

      To analyze
the contribution from the point
$t=\infty$ it is convenient  to look at the map written in terms of
Jacobi polynomials (\ref{mapjac}).
Then one can use the Rodrigues' formula to represent the Jacobi polynomials
in the form:
\be\label{JacobiExpand}
P^{(\alpha\beta)}_k(x)=2^{-k}\sum_{j=0}^k
\left(\begin{array}{c}k+\alpha\\j\\\end{array}\right)
\left(\begin{array}{c}k+\beta\\k-j\\\end{array}\right)(x-1)^{k-j}(x+1)^j.
\ee
The limit  $x\rightarrow\infty$ gives:
\be\label{JacAsympInf}
P^{(\alpha\beta)}_k(x)\rightarrow x^k 2^{-k}\sum_{j=0}^k
\left(\begin{array}{c}k+\alpha\\j\\\end{array}\right)
\left(\begin{array}{c}k+\beta\\k-j\\\end{array}\right)=
x^k 2^{-k}\left(\begin{array}{c}2k+\alpha+\beta\\k\end{array}\right).
\ee
Substitution of this limit into the expression (\ref{mapjac}) gives the
behavior near $t=\infty$ :
\bea
z&\approx& a\gamma (-1)^{d_1-d_2-n} t^{d_1-d_2},\qquad
t\approx\left((-1)^{d_1-d_2-n}\frac{z}{a\gamma}\right)^{\frac{1}{d_1-d_2}}
\nonumber\\
\label{gammafactor}
\gamma&=&\frac{d_2!(d_1-d_2-1)!}{(d_1-n)!(n-d_2-1)!}\
\frac{\Gamma(-d_1)}{\Gamma(d_2-d_1)},\\
\phi&\approx&\log\left(a\gamma(d_1-d_2)t^{d_1-d_2-1}\right)+c.c.,\qquad
\partial_t\phi\approx \frac{d_1-d_2-1}{t}.\nonumber
\eea

Consider first the  point $z=\infty, t=\infty$.  Recall that we have 
taken the `regular region' on $\Sigma$
to be bounded by the image of $1/\tilde\delta$ (rather than 
$1/\delta$) when a twist operator is
inserted.  The contour
around the puncture at infinity in the $t$  plane  should be
taken to go clockwise rather than anti-clockwise, so that it looks
like a normal anti-clockwise contour in the local coordinate
$t'=1/t$ around the puncture. Thus to compute the contribution from
this puncture we should follow our usual
procedure but reverse the overall sign.  The
result reads:
\bea
S_L(t=\infty)&=&(-1)\left[-\frac{d_1-d_2-1}{12}\log(d_1-d_2)\right.\\
&+&\left.\frac{(d_1-d_2-1)^2}{12(d_1-d_2)}\log{\tilde\delta}-
\frac{d_1-d_2-1}{12(d_1-d_2)}\log(a|\gamma|)\right].\nonumber
\eea

Finally let us analyze the  images of $z=\infty$ that give finite
values  $t_i$ of $t$. At each of these points  the map
$t\rightarrow z$ is one--to--one, in contrast to the above case
$z=\infty, t=\infty$ where $q$ values of $t$ correspond to
each value of $z$ in a neighborhood of the puncture.  Further, there
is no sign reversal  for the contour of
integration around these punctures when we use the coordinate $t$ to
describe the contour.

Looking at the structure of the map (\ref{mapjac}) one can easily identify the
locations of the $t_i$: they coincide with zeroes of the
denominator.
So  to evaluate the contribution to the Liouville action  from
the $t_i$ we will need some information about zeroes of Jacobi
polynomials. Using the fact that Jacobi polynomials have only simple zeroes we
can expand the map (\ref{mapjac}) around any of the  $t_i$:
\bea
z&\approx& at_i^n\frac{P_{d_1-n}^{(n,-d_1-d_2+n-1)}(1-2t_i)}{
{P'}_{d_2}^{(-n,-d_1-d_2+n-1)}(1-2t_i)}\ \frac{1}{t-t_i}\equiv
\frac{a\xi_i}{t-t_i},\\
\label{xi_i}
\xi_i&=&t_i^n\frac{P_{d_1-n}^{(n,-d_1-d_2+n-1)}(1-2t_i)}{
{d\over dt}{P}_{d_2}^{(-n,-d_1-d_2+n-1)}(1-2t_i)}.
\eea
Then everything can be evaluated in terms of $\xi_i$:
\bea
t-t_i&\approx& \frac{a\xi_i}{z},\qquad
\phi\approx\log\left(\frac{-a\xi_i}{(t-t_i)^2}\right)+c.c., \qquad
\partial_t\phi\approx-\frac{2}{t-t_i}\nonumber\\
S_L(t=t_i)&=&-\frac{1}{6}\log({\delta}^2a\xi_i).
\eea
Collecting the contributions from all the $t_i$ we get:
\be
S_L(\mbox{all }t_i)=-\frac{d_2}{6}\log({\delta}^2a)-
\frac{1}{6}\sum_{i=1}^{d_2}\log(\xi_i),
\ee
and we only need to evaluate the product of $\xi_i$. Note that the
regularization parameter $\delta$ we use here has the same meaning as one
considered in section \ref{Section2pt}.

This product can be written in terms of the Wronskian (\ref{WronskJacobi}) and
the discriminant of Jacobi polynomials. To see this we first rewrite
(\ref{xi_i}) in terms of zeroes of Jacobi polynomials.  If
$z=P/Q$, then the Wronskian (\ref{WronskJacobi}) is
$\tilde W=P'Q-PQ'=-PQ'$ at a zero of $Q$.  Writing any of the $\xi_i$
as $\xi=P/Q'=PQ'/Q'^2$ we find
\be
\xi_i=-{\tilde W}(t_i)\left[-2a_0\prod_{j\ne i}(x_i-x_j)\right]^{-2},
\ee
where $a_0$ is the coefficient in front of the highest power in the polynomial
$P_{d_2}^{(-n,-d_1-d_2+n-1)}$; it can be evaluated using
(\ref{JacAsympInf}).  The $x_i$ are the zeros of the polynomial
$Q(x)$ in the denominator.

Applying the general definition of the discriminant to Jacobi polynomials
\be
D_{d_2}^{(-n,-d_1-d_2+n-1)}\equiv a_0^{2d_2-2}\prod_{i< j}(x_i-x_j)^2,
\ee
we get
\be\label{prodxiStart}
\prod_{i=1}^{d_2}\xi_i=(-1)^{d_2}2^{-2d_2}a_0^{2(d_2-2)}
\left[D_{d_2}^{(-n,-d_1-d_2+n-1)}\right]^{-2}\prod_{i=1}^{d_2}{\tilde W}(t_i).
\ee
The discriminant of Jacobi polynomials can be evaluated \cite{szego}:
\bea\label{discriminant}
{\cal D}&\equiv&D_{d_2}^{(-n,-d_1-d_2+n-1)}=2^{-d_2(d_2-1)}\\
&\times&\prod_{j=1}^{d_2} j^{j+2-2d_2}
(j-n)^{j-1}(j-d_1-d_2+n-1)^{j-1}(j-d_1-1)^{d_2-j}.\nonumber
\eea
To evaluate the right hand side of (\ref{prodxiStart}) we only need the
expressions for
\be
\prod_{i=1}^{d_2} t_i \qquad \mbox{and}\qquad \prod_{i=1}^{d_2} (1-t_i).
\ee
Let us consider the general Jacobi polynomial:
\be
P_k^{(\alpha\beta)}(1-2t)=(-2)^ka_0t^k+\dots+a_{k+1}=
(-2)^kb_0(t-1)^k+\dots+b_{k+1},
\ee
Obviously $b_0=a_0$. By taking the limits $t\rightarrow \infty$,
$t\rightarrow 0$ and $t\rightarrow 1$ in the above expression we find:
\bea
\prod_{i=1}^{d_2} t_i&=&\frac{a_{k+1}}{2^k a_0}=
\frac{\Gamma(k+\alpha+1)\Gamma(k+\alpha+\beta+1)}{\Gamma(\alpha+1)
\Gamma(2k+\alpha+\beta+1)},\\
\prod_{i=1}^{d_2} (1-t_i)&=&\frac{b_{k+1}}{2^k b_0}=
\frac{\Gamma(k+\beta+1)\Gamma(k+\alpha+\beta+1)}{\Gamma(\beta+1)
\Gamma(2k+\alpha+\beta+1)}.
\eea
Collecting all contributions together, we get
\bea
\log\prod_{i=1}^{d_2}\xi_i&=&-2d_2(d_2-1)\log 2+d_2\log n-
2\log{\cal D}-(3d_2-4)\log d_2!\nonumber\\
&+&d_2\log\left[\frac{d_1!}{n!(d_1-n)!}\right]+(n+d_2-1)\log
\frac{(n-1)!}{(n-d_2-1)!}\nonumber\\
&+&(d_1-d_2+3)\log\frac{(d_1-d_2)!}{d_1!}+
(d_1+d_2-n)\log\frac{(d_1+d_2-n)!}{(d_1-n)!}.\nonumber \\
\label{ProdXi}
\eea

\subsection{The total Liouville action.}

Collecting the contributions from the different branching points we obtain the
final expression for the Liouville action
\bea\label{theLiouv3pt}
S_L^{(1)}&=&-\left(\frac{(n-1)^2}{12n}+\frac{(d_1+d_2-n)^2}{12(d_1+d_2-n+1)}
\right)
\log\epsilon-\frac{(d_1-d_2-1)^2}{12(d_1-d_2)}\log{\tilde\delta}\nonumber\\
&-&\frac{d_2}{3}\log{\delta}-
\left(\frac{n{-}1}{12n}+\frac{d_1{+}d_2{-}n}{12(d_1{+}d_2{-}n{+}1)}-
\frac{d_1{-}d_2{-}1}{12(d_1{-}d_2)}+\frac{d_2}{6}\right)\log a\nonumber\\
&-&\frac{n-1}{12}\log n-\frac{d_1+d_2-n}{12}\log(d_1+d_2-n+1)\\
&+&\frac{d_1{-}d_2{-}1}{12}\log(d_1-d_2)-
\frac{n-1}{12n}\log\left(\frac{d_1!d_2!}{n!(d_1-n)!}\
\frac{\Gamma(1{-}n)}{\Gamma(1{-}n{+}d_2)}\right)\nonumber\\
&-&\frac{d_1+d_2-n}{12(d_1+d_2-n+1)}\log|\beta|+
\frac{d_1-d_2-1}{12(d_1-d_2)}\log|\gamma|-
\frac{1}{6}\log\left(\prod_{i=1}^{d_2}\xi_i\right).\nonumber
\eea
The values of $\beta$ and $\gamma$ are given by (\ref{betafactor}) and
(\ref{gammafactor}), and the last term is given through (\ref{ProdXi}).
According to (\ref{TheCorrelator}), the three point function is given by:
\be\label{ThreePoint6.23}
\langle\sigma_{n}^\epsilon(0)
\sigma_{m}^\epsilon(a)\sigma_{q}^{\tilde\delta}(\infty)\rangle_\delta=
e^{S^{(1)}_L}e^{S^{(2)}_L+S^{(3)}_L}{Z^{(\hat s)}\over (Z_\delta)^s},
\ee
where $s$ is number of fields involved in permutation; it is defined by
(\ref{NumberOfSheets}). Note that we have not determined the values of
$S^{(2)}_L$ and $S^{(3)}_L$ for the case under consideration, as we will see
these quantities will  cancel in the final answer.

\section{Normalizing the twist operators}
\renewcommand{\theequation}{6.\arabic{equation}}
\setcounter{equation}{0}

This Liouville action (\ref{theLiouv3pt}) yields the correlation
function for twist operators with the regularization
parameters $\epsilon, \delta, \tilde \delta$. We immediately see that
the power of $a$ in the correlator is
\bea
a:&&-\left(\frac{n-1}{12n}+\frac{d_1+d_2-n}{12(d_1+d_2-n+1)}-
\frac{d_1-d_2-1}{12(d_1-d_2)}+\frac{d_2}{6}\right)\nonumber\\
&&+\frac{1}{2}\left(M_{a}^n+M_{a}^m-M_{a}^q\right)
=-\frac{1}{6}\left(n-\frac{1}{n}+m-\frac{1}{m}-q+\frac{1}{q}\right),
\eea
which agrees with the expected $a$ dependence of the 3-point function
\be
\langle
\sigma_n^\epsilon(0)\sigma_m^\epsilon(a)\sigma^{\tilde\delta}_q(\infty)\rangle
~ \sim
~|a|^{-2(\Delta_m+\Delta_n-\Delta_q)}.
\ee

      To obtain the final
correlation functions and fusion coefficients we have two sources of
renormalization coefficients that need to be considered:

(a)\quad We have to normalize the  operators $\sigma^\epsilon_n$ such
that their 2-point functions are set to
unity at unit $z$ separation; at this point we should find that the
parameters $\epsilon, \delta, \tilde \delta$ disappear from
the 3-point (and higher point) functions as well. After we normalize
the twist operators $\sigma^\epsilon_n$ in this way we
will call them $\sigma_n$.

(b)\quad The CFT had $N$ fields $X^I$, though only $n$ of them are
affected by the twist operator $\sigma^\epsilon_n$.
However at the end of the calculation of any correlation function of
the operators $\sigma^\epsilon_{n_i}$ we must sum
over all the possible ways that the $n_i$ fields that are twisted can
be chosen from the total set of $N$ fields. Thus we will
have to define operators $O_n$ that are sums over conjugacy classes
of the permutation group, and these operators $O_n$
are the only ones that will finally be well defined operators in the
CFT \cite{vafaorb}.  The correctly normalized $O_n$ will thus have
combinatoric factors multiplying the normalized operators $\sigma_n$.

We choose to arrive at the final normalized operators $O_n$ in these
two steps since the calculations involved in steps (a)
and (b) are quite different; further when $n_i<<N$ the factor coming
from (b)  is just a   power of $N$ which is easily found.

\subsection{Normalizing the $\sigma^\epsilon_n$}
\label{SecNorm}

Let us define the normalized twist operators $\sigma_n$ by requiring
\be
\langle \sigma_n(0)\sigma_n(a)\rangle~=~ {1\over |a|^{4\Delta_n}}.
\label{sfive}
\ee
     From (\ref{2ptfunction}) we see that
\be
\sigma_n=D_n\sigma^\epsilon_n, \qquad
D_n=[C_n\epsilon^{A_n}Q^{B_n}]^{-1/2}.
\label{ssix}
\ee
      Let the Operator Product Expansion (OPE) have the form
\be
\sigma_m(a)\sigma_n(0)\sim {|C^\sigma_{nmq}|^2\over
|a|^{2(\Delta_n+\Delta_m-\Delta_q)}}~\sigma_q(0)+\dots,
\label{sseven}
\ee
where we have written the OPE for holomorphic and anti-holomorphic
blocks combined, and we have put a superscript
$\sigma$ on the fusion coefficients $C_{nmq}^\sigma$ to remind
ourselves that these are not the final fusion coefficients of
the physical operators $O_n$.  With the normalization (\ref{sfive}) we will get
\bea
\langle
\sigma_n(z_1)\sigma_m(z_2)\sigma_q(z_3)\rangle=~~~~~~~~~~~~~~~~~~~~~~~
~~~~~~~~~~~~~~~~~~~~~~~~&
\nonumber\\
{|C^\sigma_{nmq}|^2\over
|z_1-z_2|^{2(\Delta_n+\Delta_m-\Delta_q)}|z_2-z_3|^{2(\Delta_m+\Delta_
q-\Delta_n)}|z_3-z_1|^{2(\Delta_q+\Delta_n-\Delta_m)}}
\label{seight}
\eea
We have computed the 3-point functions and should thus be able to get
the fusion coefficients $C^\sigma_{nmq}$ from
(\ref{seight}). However while two of our twist operators were
inserted at finite points in the $z$ plane, the last one was
inserted at infinity. Putting one of the points at infinity
simplified the calculation, but it also creates the following problem:
unlike the twist operators at $z=0, z=a$ which are normalized through
(\ref{ssix}) it is not clear what is the normalization of
the twist operator that is inserted at infinity.  (We could  think of
this operator as inserted at a puncture on the sphere at
infinity,  and therefore no different from the other insertions, but
we have chosen the flat metric on the $z$ plane and thus
made infinity a special region carrying curvature.)  To get around
this problem we adopt the following scheme. If we have
the OPE (\ref{sseven}) then we will get
\be
{\langle \sigma_n(0)\sigma_m(a)\sigma^{\tilde\delta}_q(\infty)\rangle\over
\langle \sigma_q(0)\sigma_q^{\tilde\delta}(\infty)\rangle}=
{|C^\sigma_{nmq}|^2\over
|a|^{2(\Delta_n+\Delta_m-\Delta_q)}}
\label{snine}
\ee
and we will thus not need to know the normalization of the operator at
infinity.  $\langle 
\sigma_n(0)\sigma_m(a)\sigma^{\tilde\delta}_q(\infty)\rangle$ can be 
found from
our 3-point function calculation together with the normalization
factors for $\sigma_n, \sigma_m$ from (\ref{sfive}). To
compute the denominator we must find the 2-point function with one
operator at infinity. (We had earlier computed
the 2-point function with both operators in the finite $z$ plane
since if we put one operator at infinity then we loose any
position dependence in the correlator and cannot extract the scaling
dimensions.)

To evaluate the two point function
\be
\langle \sigma_n (0)\sigma_n (\infty)\rangle
\ee
we consider the map:
\be
z=bt^n.
\ee
This map has order $n$ ramification points at $t=0$ and $t=\infty$
and the usual
calculations give:
\bea
\phi=\log[nbt^{n-1}]+c.c.,\qquad \partial_t\phi=\frac{n-1}{t},\nonumber\\
S_L(t=0)=-\frac{n-1}{12}\log[nb^{1/n}\epsilon^{(n-1)/n}],\\
S_L(t=\infty)=-(-1)\frac{n-1}{12}\log[nb^{1/n}{\tilde\delta}^{(1-n)/n}].
\eea
As before we cut a hole of size $\epsilon$ around the origin and put
the twist at infinity on a boundary at $|z|=1/{\tilde\delta}$.
We also have an extra negative sign for the cut at infinity since the
contour that goes anti-clockwise in the local
coordinate at $1/t$ near $t=\infty$  goes clockwise in the coordinate $t$.
Collecting both  contributions we get:
\bea
\langle\sigma^\epsilon_n(0)\sigma^{\tilde\delta}_n(\infty)\rangle&=&
\epsilon^{F_n}{\tilde\delta}^{F_n}
e^{S^{(2)}_L+S^{(3)}_L}{Z^{(\hat s)}\over (Z_\delta)^n},\\
F_n&=&-{(n-1)^2\over 12 n}.
\label{eone}
\eea
The correlator is $b$--independent as expected. The values of $S^{(2)}_L$,
$S^{(3)}_L$ and partition function $Z^{(\hat s)}$  depend on $\delta, 
\delta', \tilde\delta$,  but
these expressions are are the same as in (\ref{ThreePoint6.23}) and 
so they cancel in the final answer.

The $C^\sigma_{n,m,q}$ are then given by
\bea
|C^\sigma_{n,m,q}|^2&=&\frac{\langle \sigma_n(0)\sigma_m(a)
\sigma^{\tilde\delta}_q(\infty)\rangle}{\langle\sigma_q(0)
\sigma^{\tilde\delta}_q(\infty)\rangle}\\
&=&\frac{\langle \sigma^\epsilon_n(0)
\sigma^\epsilon_m(a)\sigma^{\tilde\delta}_q(\infty)\rangle}{
\langle\sigma^\epsilon_q(0)\sigma^{\tilde\delta}_q(\infty)\rangle}
\sqrt{\frac{\langle\sigma^\epsilon_q(0)\sigma^\epsilon_q(1)\rangle}{
\langle\sigma^\epsilon_n(0)\sigma^\epsilon_n(1)\rangle\langle\sigma^
\epsilon_m(0)\sigma^\epsilon_m(1)\rangle}}\ .
\nonumber
\eea
As a consistency check of our procedure we look at powers of various
regularization parameters: they should cancel in any physical quantity. For
$|C^\sigma_{n,m,q}|^2$ we have the following powers for $\epsilon,
\delta, \tilde \delta$, $Q$:
\bea
\epsilon:&&
-\left(\frac{(n-1)^2}{12n}+\frac{(m-1)^2}{12m}\right)-F_q-
\frac{1}{2}\left(A_n+A_m-A_q\right)=0,\\
{\tilde\delta}:&& -\frac{(q-1)^2}{12q}-F_q=0,\\
\delta:&& -\frac{d_2}{3}-s+q=0,\\
Q:&&1-s-\frac{1}{2}\left(B_n+B_m+B_q\right)=0.
\eea
We used the expressions (\ref{expr_d1}) and (\ref{expr_d2}) for $d_1$ and
$d_2$, the values of $A_n, B_n, F_n$ from (\ref{2ptfuncoef}) and
(\ref{eone}) and the genus relation (\ref{GenusG=0}).

We finally get   (for the contribution from $\Sigma$ of genus zero) 
the logarithm of the fusion
coefficient
\bea
\log| C^\sigma_{n,m,q}|^2&=&\frac{1}{6}\log\left(\frac{q}{mn}\right)
-\frac{n-1}{12}\log n-\frac{m-1}{12}\log m +\frac{q-1}{12}\log(q)
\nonumber\\
&-&\frac{n-1}{12n}\log\left(\frac{d_1!d_2!}{n!(n-1)!}\
\frac{(d_1-m)!}{(d_1-n)!}\right)\\
&-&\frac{m-1}{12m}\log\left(\frac{d_1!d_2!}{m!(m-1)!}\
\frac{(d_1-n)!}{(d_1-m)!}\right)\nonumber\\
&+&\frac{q{-}1}{12q}\log\left(\frac{(q{-}1)!d_2!}{(d_1{-}n)!(d_1-m)!}\
\frac{(d_1-d_2)!)}{d_1!}\right)-
\frac{1}{6}\log\left(\prod_{i=1}^{d_2}\xi_i\right).\nonumber
\label{etwo}
\eea
The expression for the product of $\xi_i$ is given by (\ref{ProdXi}).

   The coefficients  $C^\sigma_{n,m,q}$ must be symmetric in the indices
$m,n,q$. We have written (\ref{etwo}) in such a way
that all the terms except the last one show a manifest symmetry
between $m$ and $n$. It can be shown without much difficulty
that the last term (given through (\ref{ProdXi}) and (\ref{discriminant}))
is also symmetric in $m$ and $n$. In
particular $d_1$ and $d_2$ are symmetric  in $m$ and $n$, and
$-d_1-d_2+n-1=-m$, so that the Jacobi polynomial whose
discriminant is calculated in  (\ref{discriminant}) is $P^{-n, -m}_{d_2}$.

On the other hand it is not at all obvious that the expression
(\ref{ProdXi}) is symmetric under the interchange of $q$ with
either $n$ or $m$. Note that $d_2+1$ is the number of elements that
overlap between the permutations $\sigma_n$ and
$\sigma_m$, and the product in  (\ref{discriminant}) runs over the
range $j=1\dots d_2$. This number $d_2+1$ is in general
different from the number of overlapping elements between the
permutations $\sigma_q$ and
$\sigma_n$ or between $\sigma_q$ and
$\sigma_m$, and thus there is no simple way to write (\ref{ProdXi})
in form that makes its total symmetry manifest.
Nevertheless, this expression is indeed symmetric in all three
arguments $n,m,q$, as can be checked by evaluating the
expression through a symbolic manipulation program. Verifying this
symmetry provides a useful check of all our
calculations for the 3-point function.

\subsection{Two special cases}

Due to the structure of the discriminant (\ref{discriminant}), the general
expression for the fusion coefficient looks complicated for an arbitrary
value of $d_2$. However there are two important cases where significant
simplifications occur. These are the cases of one and two overlaps (we recall
that the number of common indices in $\sigma_n$ and $\sigma_m$ is $d_2+1$).
One can see that for $d_2=0$ and $d_2=1$ the discriminant ${\cal D}=1$. Let
us analyze both these cases.

For one overlap we have:
\be
d_2=0,\qquad d_1=q=m+n-1,
\ee
and the logarithm of fusion coefficient is given by:
\begin{eqnarray}\label{result1over}
&&\log |C^\sigma_{n,m,m+n-1}|^2=-\frac{1}{12}\left(n+\frac{1}{n}\right)\log n-
\frac{1}{12}\left(m+\frac{1}{m}\right)\log m\\
&&+\frac{1}{12}\left(q{+}\frac{1}{n}{+}\frac{1}{m}-1\right)\log q-
\frac{1}{12}\left(1{+}\frac{1}{q} {-}\frac{1}{n} {-}\frac{1}{m}\right)
\log\left(\frac{(q{-}1)!}{(m{-}1)!(n{-}1)!}\right).\nonumber
\eea
In particular we get:
\be\label{C223}
|C^\sigma_{223}|^2=2^{-\frac{4}{9}}3^{\frac{1}{4}}.
\ee
The case of two overlaps corresponds to
\be
d_2=1,\qquad q=m+n-3,\qquad d_1=q+1,
\ee
and the result reads:
\bea
\log
|C^\sigma_{n,m,m{+}n{-}3}|^2=\frac{1}{12}\left(\frac{1}{n}{+}\frac{1}{
m}{-}
\frac{1}{m{+}n{-}3}-3\right)\log\left(\frac{(m{+}n{-}3)!}{(m{-}1)!(n{-}1)!}
\right)\nonumber\\
-\frac{n^2+1}{12n}
\log n-
\frac{m^2+1}{12m}\log m+
\frac{1}{12}\left(2{+}\frac{(m{+}n{-}4)^2}{m{+}n{-}3}\right)\log(m{+}n{-}3)+
\nonumber
\eea
$$
\frac{1}{12}\left(\frac{n{-}m}{mn} {-}2n{+}
\frac{m{+}n{-}4}{m{+}n{-}3}\right)
\log(n{-}1)+
\frac{1}{12}\left(\frac{m{-}n}{mn} {-}2m{+}\frac{m{+}n{-}4}{
m{+}n{-}3}\right)
\log(m{-}1)\nonumber
$$
\be\label{result2over}
+\frac{1}{12}\left(2(m+n)-5+\frac{1}{n}+\frac{1}{m}+\frac{1}{m+n-3}\right)
\log(m+n-2).
\ee
In particular for $m=3$ we get:
\bea\label{Cn3n}
\log |C^\sigma_{n3n}|^2&=&-\frac{2}{9}\log n-
\frac{1}{6}\left(n+\frac{1}{n}-\frac{2}{3}\right)\log(n-1)\nonumber\\
&+&\frac{1}{6}\left(n+\frac{1}{n}+\frac{2}{3}\right)\log(n+1)-
\frac{2}{9}\log 2-\frac{5}{18}\log 3.
\eea
  From this expression we can extract the value of $C^\sigma_{232}$ and
check that it
equals the value of  $C^\sigma_{223}$ given by (\ref{C223}).

\subsection{Combinatoric factors and large N limit.}

The twist operators we have considered so far do not represent proper fields
in the conformal field theory. In the orbifold CFT there is one twist field
for each conjugacy class of the permutation group, not for each
element of the group \cite{vafaorb}. The true CFT
operators that represent the twist fields can be constructed by
summing over the group orbit:
\be
O_n=\frac{\lambda_n}{N!}\sum_{h\in G} \sigma_{h(1\dots n)h^{-1}}\ .
\ee
Here $G$ is the permutation group $S_N$ and the normalization constant
${\lambda_n}$ will be determined below.

Using normalization condition for the $\sigma$ operators:
\be
\langle\sigma_n(0)\sigma_n(1)\rangle=1
\ee
we find:
\be
\langle O_n(0)O_n(1)\rangle={\lambda_n^2\over N!}\langle\sigma_{(1\dots n)}
\sum  \sigma_{h(1\dots n)h^{-1}}\rangle=\lambda_n^2n\frac{(N-n)!}{N!}
\langle\sigma_n(0)\sigma_n(1)\rangle.
\ee
Requiring the normalisation $\langle O_n(0)O_n(1)\rangle=1$ we find
the value of
$\lambda_n$:
\be
\lambda_n=\left[{n(N-n)!\over N!}\right]^{-1/2}.
\ee
Let us now look at the three point function. First we  consider the
combinatorics for the $g=0$ cases that we worked
with above; the permutation structure was described in (\ref{thone}),
(\ref{thtwo}), (\ref{ththree}).  Simple
combinatorics yields
\bea
\langle O_n(0)O_m(1)O_q(z)\rangle=({1\over N!})^3\lambda_n\lambda_m
\lambda_q~~~~~~~~~~~~~~~~~~~~~~~~~~~~~\nonumber \\
\times nmq{N!\over (N-s)!}(N-n)!(N-m)!(N-q)!
\langle\sigma_n(0)\sigma_m(1)\sigma_q(z)\rangle.
\label{ethree}
\eea
One way of getting this expression is to note that $s$ different
indices are involved in the permutation, and we can
select these indices, in the order in which they appear when the
permutations are written out, in $N!/(N-s)!$ ways.
Having obtained the indices for any given permutation, we ask how many
elements out of the sum over group elements
yields this set of indices in the permutation; the answer for
$\sigma_n$ for example is $(N-n)!$, since only the
permutations of the remaining $N-n$ elements leave the indices in
$\sigma_n$ untouched. Finally, we note that any
permutation $\sigma_k$ can be written in $k$ equivalent ways since
we can begin the set of indices with any index
that we choose from the set; this leads to the factors $nmq$.

      Substituting the values of $\lambda_i$ we get the final result:
\be\label{combinatorics}
\langle O_n(0)O_m(1)O_q(z)\rangle{=}
\frac{\sqrt{mnq(N-n)!(N-m)!(N-q)!}}{(N-s)!
\sqrt{N!}}
\langle\sigma_n(0)\sigma_m(1)\sigma_q(z)\rangle.\nonumber
\ee
with $s={1\over 2}(n+m+q-1)$.

Now we analyse the behavior of the combinatoric factors  for
arbitrary genus $g$ but in the limit where
      $N$ is taken to be large  while the orders of twist
operators ($m$, $n$
and $q$) as well as the  parameter $g$ are kept fixed.
There are $s$ different fields $X^i$ involved in the 3-point
function, and these fields can be selected in $\sim N^s$ ways.
Similarily the 2-point function of $\sigma_n$ will go as $N^n$ since
$n$ different fields are  to be selected. Thus the 3-point
function of normalised twist operators will behave as
\be
N^{s-{n+m+q\over 2}}= N^{-(g+{1\over 2})}
\label{efour}
\ee
(which can also be obtained from (\ref{combinatorics})).

Thus in the large $N$ limit the contributions from surfaces with high genus
will be suppressed, and  the leading order the answer can be obtained by
considering only contributions from the sphere ($g=0$). This is presicely the
case that we have analysed in detail, and knowing the amplitude
$\langle\sigma_n(0)\sigma_m(1)\sigma_q(z)\rangle$ one can easily extract the
leading order of the CFT correlation function:
\be
\langle O_n(0)O_m(1)O_q(z)\rangle=\sqrt{\frac{1}{N}}
{\sqrt{mnq}}\langle\sigma_n(0)\sigma_m(1)\sigma_q(z)
\rangle_{sphere}+O\left(\frac{1}{N^{3/2}}\right).
\ee

\section{Four Point Function.}
\renewcommand{\theequation}{7.\arabic{equation}}
\setcounter{equation}{0}

In this section we compute specific examples of 4-point functions,
without attempting to analyze the most general case.
The computations illustrate interesting features which
arise in our approach for four and higher point
functions. In particular we will also need to compute a genus
one correlation function. We will also be able to verify specific
examples of the fusion coefficients computed in the last section as
they will be recovered through factorization of the
4-point functions.

\subsection{An example of a 4-point function on a sphere.}
\label{fourpoint}

Let us start with a map that has branch points appropriate for a
4-point correlation function of the form
\be
\langle \sigma_n(0) \sigma_2(1)\sigma_2(w)\sigma_n(\infty)\rangle
\label{enine}
\ee
Consider the map
\be\label{4ptmap1}
z=Ct^n\frac{t-a}{t-1},
\ee
where the parameter $a$ will be related with coordinate $w$ and the value of
coefficient $C$ will be determined below. The map (\ref{4ptmap1}) has two
obvious ramification points: $z=0$ and $z=\infty$, both of them give
an  $n$--th
order branch point for nonzero values of $a$. For a general value of
$a$ the map
(\ref{4ptmap1}) has two more ramification points; to find them we should look
at the equation
\be
\frac{dz}{dt}=0.
\ee
For general value of $a$ this equation reads:
\be
t^{n-1}(nt^2-t((n-1)a+(n+1))+an)=0.
\label{4ptQuadrat}
\ee
The first factor corresponds to the obvious fact that at the point $t=0$ we
have a ramification point of $n$--th order, while the positions of the two
``implicit'' points of second order are given by:
\be
t_\pm=\frac{1}{2n}\left((n-1)a+n+1\pm\sqrt{(a-1)((n-1)^2a-(n+1)^2)}\right).
\ee
One of these points should correspond to $z=1$; we let this be the
point $t_+$.  The other must  correspond to
$z=w$.  By requiring $z(t_+)=1$ we determine the value of coefficient $C$:
\be\label{4ptCoeffc}
C=t_+^{-n}\frac{t_+-1}{t_+-a},
\ee
       and we note that in what follows we will have
\be
w=t_-.
\ee
Now we will analyze contributions to the Liouville action coming from
the different
      ramification points. Let us start from the point $t=0$. If $a\ne 0$
the map (\ref{4ptmap1}) near this point has the form:
\be
z\approx Cat^n
\ee
and the inverse map is
\be
t\approx \left(\frac{z}{aC}\right)^{1/n}.
\ee
The Liouville field and its derivative are given by:
\be
\phi=\log\left(\frac{dz}{dt}\right)+c.c.\approx\log(nCat^{n-1})+c.c., \qquad
\partial_t\phi\approx\frac{n-1}{t}.
\ee
As usual we will cut a hole of radius $\epsilon$ around the point $z=0$.
Then the contribution to the Liouville action coming from the integration
over the boundary of this hole is
\be
S_L(t=0)=-\frac{n-1}{12}\log\left[n(aC)^{1/n}\epsilon^{\frac{n-1}{n}}\right].
\ee
The same analysis near $t=\infty$ gives:
\bea
z\approx Ct^n, \qquad t\approx \left(\frac{z}{C}\right)^{1/n},\nonumber\\
\phi\approx\log(nCt^{n-1})+c.c., \qquad
\partial_t\phi\approx\frac{n-1}{t},\nonumber\\
S_L(t=\infty)=-(-1)\frac{n-1}{12}
\log\left[nC^{1/n}{\tilde\delta}^{\frac{1-n}{n}}\right].
\eea
Here we have cut a large circle of radius $1/{\tilde\delta}$ in $z$ plane and
the factor of
$(-1)$ in the last equation comes from the fact that we go around the point
$t=\infty$ clockwise.

Near the point $t=t_-$ we get:
\bea
z\approx z_-+\xi_-(t-t_-)^2, \qquad
t-t_-\approx\left(\frac{z}{\xi_-}\right)^{1/2}, \nonumber\\
\xi_-=\frac{1}{2}\left(\frac{d^2z}{dt^2}\right)_{t=t_-}\approx
\frac{nz_-(t_--t_+)}{2t_-(t_--a)(t_--1)},\nonumber\\
\phi\approx\frac{1}{2}\log(4(z-z_-)\xi_-)+c.c., \qquad
\partial_t\phi\approx\frac{1}{t},\nonumber\\
S_L(t=t_-)=-\frac{1}{24}
\log\left[4\epsilon\xi_-\right].
\eea
To get a contribution for $t=t_+$ one should make a replacement
$+\leftrightarrow -$ in the last expression; we also note that $z_+=1$. Thus
we get:
\bea
S_L(t=t_+)=-\frac{1}{24}
\log\left[4\epsilon\xi_+\right],\\
\xi_+\approx
\frac{n(t_+-t_-)}{2t_+(t_+-a)(t_+-1)}.\nonumber
\eea
Finally we should consider the  images of $z=\infty$ that give finite
values for $t$ - there will be a puncture here
corresponding to the boundary of the  $|z|$ plane. As before we let
this circle in the $z$ plane have a radius $1/\delta$.
   The map  (\ref{4ptmap1}) has only one such image:
$t=1$. The result is
\bea
z\approx C\frac{1-a}{t-1}, \qquad t-1\approx\frac{C(1-a)}{z},
\nonumber\\
\phi\approx\log\left(-\frac{z^2}{C(1-a)}\right)+c.c., \qquad
\partial_t\phi\approx-\frac{2}{t-1}\nonumber\\
S_L(t=1)=-\frac{1}{6}
\log\left[C(1-a){ {\delta}}^2\right].
\eea
Collecting all this information together and making obvious simplifications we
finally get the expression for the Liouville action corresponding to our four
point function:
\bea
S_L^{(1)}&=&\left(-\frac{(n-1)^2}{12n}-\frac{1}{12}\right)\log\epsilon-
\frac{(n-1)^2}{12n}\log{\tilde\delta}-\frac{1}{3}\log{\delta}\nonumber\\
&-&\frac{1}{4}\log C-\frac{1}{12}\log 2-
\frac{1}{12}\left(\frac{n+1}{2}-\frac{1}{n}\right)\log a-\frac{1}{8}\log(1-a)
\nonumber\\
&-&\frac{1}{24}\log\left[(n-1)^2 a-(n+1)^2\right]-\frac{1}{12}\log n.
\eea

The above expression gives the correlator $\langle
\sigma^\epsilon_n(0)\sigma^\epsilon_2(1)\sigma^\epsilon_2(w)\sigma^{
\tilde\delta}_n
(\infty)\rangle$.
To compute the correlation function of normalised twist operators, 
with one point at
infinity, defined as
\bea
\langle
\sigma_n(0)\sigma_2(1)\sigma_2(w)\sigma_n(\infty)\rangle&\equiv&
\lim_{|z|\rightarrow\infty}|z|^{4\Delta_n}\langle\sigma_n(0)\sigma_2(1
)\sigma_2(w)\sigma_n(z)\rangle\nonumber\\
&=&{\langle
\sigma_n(0)\sigma_2(1)\sigma_2(w)\sigma^{\tilde\delta}_n(\infty)\rangle\over
\langle\sigma_n(0)\sigma_n^{\tilde\delta}(\infty)\rangle}
\eea
we use arguments similar to those in subsection \ref{SecNorm}. Then we find
\bea
{\cal
F}_4&\equiv&\langle\sigma_n(0)\sigma_2(1)\sigma_2(w)\sigma_n(\infty)
\rangle\nonumber \\
&=&{\langle
\sigma^\epsilon_n(0)\sigma^\epsilon_2(1)\sigma^\epsilon_2(w)\sigma^
{\tilde\delta}_n(\infty)\rangle\over
\langle\sigma^\epsilon_n(0)\sigma_n^{\tilde\delta}(\infty)\rangle}[\langle
\sigma_2^\epsilon(0)\sigma^\epsilon_2(1)\rangle]^{-1}
\eea
This leads to
\bea
\log{\cal F}_4&=&S_L^{(1)}-
\log\left(\epsilon^{F_n}{\tilde\delta}^{F_n}\right)-
\log\left(C_2\epsilon^{A_2}Q^{B_2}\right 
)+\frac{1}{3}(n+1-n)\log\delta\nonumber \\
&+&((n-1)-n)\log Q.
\eea
where  $A_n, C_n,  F_n$ are given in (\ref{2ptfuncoef}) and (\ref{eone}). The
sourse of the last two terms is the fact that the numerator has $n+1$ fields
transforming nontrivially, while the denominator has only $n$:
\bea
\langle\sigma^\epsilon_n(0)\sigma^{\tilde\delta}_n(\infty)\rangle&=&
\epsilon^{F_n}{\tilde\delta}^{F_n}
e^{S^{(2)}_L+S^{(3)}_L}{Z^{(\hat s)}\over (Z_\delta)^n},\\
\langle
\sigma^\epsilon_n(0)\sigma^\epsilon_2(1)\sigma^\epsilon_2(w)\sigma^
\delta_n(\infty)\rangle&=&
e^{S^{(1)}_L}e^{S^{(2)}_L+S^{(3)}_L}{Z^{(\hat s)}\over (Z_\delta)^{n+1}}
\eea
We observe that the powers of regularization parameters
cancel:
\bea
\log\epsilon:&& -\frac{(n-1)^2}{12n}-\frac{1}{12}-F_n-A_2=0,
\nonumber\\
\log{\tilde\delta}:&& -\frac{(n-1)^2}{12n}-F_n=0,\qquad
\log{\delta}:\ \ -\frac{1}{3}+\frac{1}{3}=0,\\
\log Q:&&-n+(n-1)+(2-1)=0.\nonumber
\eea
This  cancellation gives a consistency check on the  calculations.
The expression for the
logarithm of the normalized four point function is given by:
\bea
\log{\cal F}_4&=&-\frac{1}{4}\log C-
\frac{1}{12}\left(\frac{n+1}{2}-\frac{1}{n}\right)\log a-\frac{1}{8}\log(1-a)
\nonumber\\
\label{result4ptFunc}
&-&\frac{1}{24}\log\left[(n-1)^2 a-(n+1)^2\right]-
\frac{1}{12}\log n-\frac{5}{12}\log 2.
\eea
The value of $C$ is given by (\ref{4ptCoeffc}).

\subsection{Analysis of the 4-point function}

Let us step back from the above calculation and think about the
structure of a 4-point function
$\langle\sigma_n(0)\sigma_2(1)\sigma_2(w)\sigma_n(\infty)\rangle$.
Consider the limit $w\rightarrow 0$, and ask what
operators are produced in the OPE of $\sigma_n$ and $\sigma_2$. There
are three possibilities, which must all be
considered when we  make the CFT operators $O_j$ out of the sum over
indices in the $\sigma_j$:

(a)\quad The indices of $\sigma_2$ and the indices of $\sigma_n$ have
no overlap - i.e., the operators are of the form
$\sigma_{12}\sigma_{34\dots n+2}$. In this case the other two
operators must have no overlapping indices either, and the
entire 4-point function factors into two different parts
$\langle\sigma_2\sigma_2\rangle\langle\sigma_n\sigma_n\rangle$.
The covering surfaces are separate for the two parts, and we just
multiply together  the correlation functions obtained
from the covering surfaces for the 2-point functions.

(b)\quad The indices of $\sigma_2$ and the indices of $\sigma_n$ have
one overlap - i.e., the operators are of the form
$\sigma_{12}\sigma_{23\dots n+1}$.  The OPE then produces the
operator $\sigma_{123\dots n+1}=\sigma_{n+1}$. The other
two operators in the correlator must also have a singe overlap so
that they can produce $\sigma_{n+1}$. The genus of the
surface thus produced is seen to be
\be
g={1+1+(n-1)+(n-1)\over 2}-(n+1)+1=0
\ee
This case in fact corresponds to the surface that was constructed in
the subsection above. Note that if we take
$\sigma_{12}$ around $\sigma_{23\dots n+1}$ then it becomes
$\sigma_{13}$.  The OPE of $\sigma_{13}$ with
$\sigma_{23\dots n+1}$ is still an operator  of the form
$\sigma_{n+1}$. On the other hand if we take the two $\sigma_2$
operators near each other, then we get the identity if we have
$\sigma_{12}\sigma_{12}$, but we get $\sigma_{123}$ if
move the operators through a  path such that they become
$\sigma_{12}$ and $\sigma_{13}$. In fact by moving the various
operators around each other on the $z$ plane, we can also get from
the same correlator OPEs of the form
$\sigma_{12}\sigma_{34}$ (which is nonsingular) and
$\sigma_{12}\sigma_{n,n-1,\dots 21}$ which produces an operator
$\sigma_{n-1}$. Thus we should find singularities in the 4-point
function arising from this surface to correspond to all
these possibilities.

(c)\quad The indices of $\sigma_2$ and the indices of $\sigma_n$ have
two overlaps, and the total number of indices
involved in the correlator is $s=n$. (Note that case (b) above also
could be brought to a form where $\sigma_2$ and
$\sigma_n$  have two overlaps,  but the number of indices involved there
overall was $n+1$.) The other two operators in the correlator must
have a similar
      overlap of indices, since otherwise they cannot produce an operator
that has only $n$ distinct indices.  In this case the
genus of the covering surface is
\be
g={1+1+(n-1)+(n-1)\over 2}-n+1=1.
\ee

We see that the correlator $\langle O_2O_2O_2O_2\rangle$ will have
contributions from correlators $\langle
\sigma_2\sigma_2\sigma_2\sigma_2\rangle$ that give genus 0 and genus
1 surfaces, but no other surfaces. The
genus 0 case is contained in the analysis in subsection
\ref{fourpoint}, and we will study the genus 1 case in
subsection
\ref{SubSectG1} below. Note that for the genus 0 case we have many
combinations of indices for the $\sigma_2$
operators as discussed in (b) above, but these all arise from
different branches of the same function
(\ref{result4ptFunc}). We must thus add the results from these
branches (as well as the disconnected part (case
(a) above) and the genus 1 contribution) to obtain the complete
4-point function of the $O_2$ operators. We will
not carry out the explicit addition since we expect the result to be
simpler in the supersymmetric case, which we
hope to present elsewhere.

\subsection{Analysis of the $g=0$ contribution}

In this subsection, we analyse the correlator computed in
(\ref{result4ptFunc}), which
corresponds to case (b) above, to check if it reproduces the expected
short distance limits.

\subsubsection{The limit $w\rightarrow 0$.}

First let us consider the limit $w\rightarrow 0$, which
corresponds to $t_-\rightarrow 0$ and $a\rightarrow 0$. For small values of
$a$ we have:
\bea
t_+&\approx&\frac{n+1}{n}, \qquad t_-\approx\frac{an}{n+1}, \qquad
C\approx \frac{n^n}{(n+1)^{n+1}},\nonumber\\
z_-&\approx& n^{2n}(n+1)^{-2-2n}a^{n+1},\nonumber\\
\log{\cal F}_4&\approx&\frac{1}{12}\left(\frac{1}{n(n+1)}-\frac{1}{2}\right)
\log z_--\frac{5}{12}\log 2\\
&-&\left(\frac{n}{6}+\frac{1}{6(n+1)}+\frac{1}{12}\right)\log n+
\left(\frac{n}{6}+\frac{1}{6n}+\frac{1}{12}\right)\log(n+1).\nonumber
\eea
One can see that the correct singularity $(z_-)^{ -2(\Delta_n+\Delta_2-
\Delta_{n+1})}$ is reproduced. Using the expressions for three point functions
      we derived before one can check that in the limit $w\approx 0$:
\be
\log{\cal F}_4\approx -2(\Delta_n+\Delta_2-\Delta_{n+1})\log w+
2\log |C^\sigma_{n,2,n+1}|^2,
\ee
which agrees with anticipated factorization.

\subsubsection{The limit $w\rightarrow 1$}
Let us now consider a limit $a\rightarrow 1$, which corresponds to one of two
possible ways for point $w$ to approach $1$. After introducing $b=a-1$ we get:
\bea
t_\pm&\approx& 1+\frac{b(n-1)}{2n}\pm\frac{1}{2n}\sqrt{-4nb},\qquad C\approx 1,
\nonumber\\
t_\pm-a&\approx& \pm\sqrt{-\frac{b}{n}}-\frac{b}{n}\frac{n+1}{2},\nonumber\\
z_--z_+&=&\frac{z_-}{z_+}-1\approx\left(1-2\sqrt{-\frac{b}{n}}\right)^n
\left(1-(n-1)\sqrt{-\frac{b}{n}}\right)\nonumber\\
&\times&\left(1-(n+1)\sqrt{-\frac{b}{n}}\right)-1\approx -4i\sqrt{nb},
\nonumber\\
\log{\cal F}_4&\approx&-\frac{1}{4}\log(w-1)=-4\Delta_2\log(w-1).
\eea
This singularity corresponds to $\sigma_2$ and $\sigma_2$ fusing to 
the identity.

There is another limit ($a\rightarrow
\frac{(n+1)^2}{(n-1)^2}$) which also corresponds to $w\rightarrow 1$.
Introducing $b=a-(n+1)^2/(n-1)^2$, we get:
\bea
t_\pm&\approx& \frac{n+1}{n-1}\pm\sqrt{\frac{b}{n}},\qquad
C\approx -\left(\frac{n-1}{n+1}\right)^{n+1},
\nonumber\\
\frac{dz}{dt}&=&\frac{Cnt^n}{(t-1)^2}(t-t_+)(t-t_-)\approx
-\frac{n(n-1)^4}{4(n+1)^2}(t-t_+)(t-t_-)\nonumber\\
z_--z_+&\approx&-\frac{n(n-1)^4}{4(n+1)^2}\int_{t_+}^{t_-}dt (t-t_+)(t-t_-)=
-\frac{(n-1)^4b^{3/2}}{3\sqrt{n}(n+1)^2},\nonumber\\
\log{\cal F}_4&\approx&-\frac{1}{36}\log(w-1)+
\log(n-1)\left[\frac{1}{9}-\frac{1}{6}\left(n+\frac{1}{n}\right)\right]-
\frac{2}{9}\log n\nonumber\\
&+& \log(n+1)\left[\frac{1}{6}\left(n+1+\frac{1}{n}\right)-\frac{1}{18}\right]-
\frac{2}{3}\log 2-\frac{1}{36}\log 3.\nonumber\\
& &
\eea
Using the equations (\ref{Cn3n}) and (\ref{C223}) one can see that
\be
\log{\cal F}_4\approx-2(\Delta_2+\Delta_2-\Delta_3)\log(w-1)+\log 
|C^\sigma_{223}|^2+
\log |C^\sigma_{n3n}|^2.
\ee
This corresponds to merging $\sigma_2$ and $\sigma_2$ to $\sigma_3$.

\subsubsection{The limit $w\rightarrow \infty$.}

We now look at the remaining limits of the expression (\ref{result4ptFunc})
at which the four point function becomes singular. They emerge at the points
where the coefficient $C$ goes either to $0$ or infinity, i.e. if the value of
$t_+$ approaches one of the points: $0,\ 1,\ a,\ \infty$. Substituting this to
the quadratic equation (\ref{4ptQuadrat}), we get the candidates for the
critical values of $a$: $0,\ 1,\ \infty$. Two of these limits we already
considered, now we analyze the last possibility: $a\rightarrow\infty$. In
this limit we have:
\be
t_+\approx a\frac{n-1}{n}-\frac{1}{n(n-1)},\quad
t_-\approx\frac{n}{n-1}.
\ee
So it is convenient to keep a point $z_-$ fixed and vary the value of $z_+$
instead\footnote{Note that the definition of $t_+$ and $t_-$  depend
on the choice of branch  for a
multivalued function; in particular $t_+$ and $t_-$
interchange if one goes along a small circle around the point $a=1$.}. Thus we
get:
\bea
C&=&t_-^{-n}\frac{t_--1}{t_--a}\approx -\left(\frac{n-1}{n}\right)^{n}
\frac{1}{a(n-1)},
\nonumber\\
z_+&\approx&a^{n-1}\left(\frac{n-1}{n}\right)^{2n}\frac{1}{(n-1)^2},\nonumber\\
\log{\cal F}_4&\approx&\left(-\frac{1}{24}+\frac{1}{12n(n-1)}\right)\log z_+ +
\log(n-1)\left[\frac{1}{12}-\frac{1}{6}\left(n+\frac{1}{n}\right)\right]
\nonumber\\
&+&\left[\frac{1}{6}\left(n-1+\frac{1}{n-1}\right)+\frac{1}{12}\right]\log n-
\frac{5}{12}\log 2.
\eea
This expression can be rewritten in terms of three point functions:
\be
\log{\cal F}_4\approx 2(\Delta_n-\Delta_2-\Delta_{n-1})+2\log 
|C^\sigma_{n-1,2,n}|^2,
\ee
thus it corresponds to the factorization of the following type:
\be
\left\langle \left(\sigma_{(1\dots n)}\sigma_{(12)}\right)
\left(\sigma_{(12)}\sigma_{(1\dots n)}\right)\right\rangle.
\ee

Thus the four point function reproduces the anticipated factorizations.

\subsection{The $g=1$ correlator $\langle
\sigma_{12}\sigma_{12}\sigma_{12}\sigma_{12}\rangle$}
\label{SubSectG1}
Let us consider the case $n=2$ in (\ref{enine}), so that we have the correlator
\be
\langle \sigma_2(0) \sigma_2(1)\sigma_2(w)\sigma_2(\infty)\rangle
\ee
We wish to have the number of sheets over a generic point in the $z$
plane to be $2$; this gives $g=1$ for the covering
surface $\Sigma$.  Each branch point is of order $2$, so we seek a
map of the form
\be
{dz\over dt}=\alpha[z(z-1)(z-w)(z-z_\infty)]^{1/2}
\label{none}
\ee
We choose not to put any branch point at infinity explicitly, the limit
$z_\infty\rightarrow\infty$ will be taken in the end of the calculation.
This equation may be solved using the Weierstrass function ${\cal P}$
and the solution in the $z_\infty\rightarrow\infty$ limit is given by:
\be
z(t)={{\cal P}(t)-e_1\over e_2-e_1}
\ee
where
\bea
e_1&=&{\cal P}({1\over 2}), ~~~e_2={\cal P}({\tau\over 2}),
~~~e_3={\cal P}({1\over 2}+{\tau\over 2})\nonumber \\
w&=&\left({\theta_3(\tau)\over \theta_4(\tau)}\right)^4={e_3-e_1\over e_2-e_1}
\label{ntwo}
\eea
The coordinate $t$ describes a torus given by modding out the complex
plane with translations by $1$ and $\tau$.

We choose the fiducial metric on the torus to be that flat metric
$d\hat s^2=dtd\bar t$. Then we calculate the
contribution to the Liouville action from the point $z=0$. Near this point
\bea
{dz\over dt}&\approx& \alpha z^{1/2} \sqrt{-wz_\infty}\nonumber \\
\phi&=&\log {dz\over dt}+c.c.~\approx~\log\left(
\alpha z^{1/2} \sqrt{-wz_\infty}\right)+c.c.\nonumber \\
\partial_t\phi&=&{dz\over dt}\partial_z\phi ,
~~~~~\partial_z\phi\approx {1	\over 2z}\nonumber \\
\label{nthree}
\eea
We can write $dt\partial_t\phi$=$dz\partial_z\phi$ for any
infinitesimal segment of the contour of integration around the
puncture, but we must circle the $z$ plane puncture twice to circle
the $t$ space puncture once.  We find it easier to work
in the $z$ plane instead of the $t$ space to evaluate the integral
around the puncture. We recall that the puncture has a
radius $\epsilon$ in the $z$ plane,  and we put in a factor of
$2$ at the end to account for the relation between $z$ and $t$
contours. Then we get for the contribution to $S_L$:
\be
S_L(z=0)=-{1\over 24}\log\left[\alpha^2\epsilon |wz_\infty|\right]
\ee
Similarly from the points $z=1$, $z=w$ and $z=z_\infty$ we get the
contributions
\bea
S_L(z=1)=-{1\over 24}\log\left[\alpha^2\epsilon
|1-w||1-z_\infty|\right],\\
S_L(z=w)=-{1\over 24}\log\left[\alpha^2\epsilon |w||1-w||w-z_\infty|\right],\\
S_L(z=z_\infty)=-{1\over 24}\log\left[\alpha^2\epsilon |z_\infty||z_\infty|
|w-z_\infty|\right].
\eea
Now we look at the image of infinity, where we have cut a
circle $|z|=1/\delta$. We have
\be
\frac{dz}{dt}\approx\alpha z^2, \qquad
\phi~\approx~\log[\alpha z^2]+c.c.,\qquad
\partial_z\phi\approx{2\over z}
\ee
Noting that we must take the $z$ plane contour clockwise, and putting
in the factor of $2$ to relate the contour to the $t$
space contour, we get the contribution
\be
S_L(z=\infty)=\frac{1}{3}\log\left[\alpha\delta^2\right]
\ee
Adding up all contributions we get
\bea
S_L^{(1)}&=&-{1\over 6}\log\epsilon-{2\over 3}\log \delta-{1\over 12}
\log\left|w(1-w)z_\infty (z_\infty-1)(z_\infty-w)\right|\nonumber\\
&\approx&-{1\over 6}\log\epsilon-{2\over 3}\log \delta-
{1\over 12}\log\left|w(1-w)\right|-\frac{1}{4}\log |z_\infty|
\eea
Note that $\alpha$ does not appear in the final result, as one could
anticipate from the fact that this constant can be absorbed in the rescaling
of the $t$ plane. In the case under consideration there is no contribution to
the Liouville action coming from the $|z|>1/\delta$ region: there is no
curvature on the torus $t$ and
\be
\frac{d{\tilde z}}{dt}=-\frac{1}{\delta^2 z^2}\frac{dz}{dt}\approx
-\frac{\alpha}{\delta^2},
\ee
giving a constant $\phi$ to leading order and thus a vanishing
kinetic term for $\phi$.
Thus $S_L=S_L^{(1)}$ and the general expression (\ref{TheCorrelator}) gives:
\be
\langle \sigma_2(0)\sigma_2(1)\sigma_2(w)\sigma_2(z_\infty)\rangle_\delta=
e^{S_L^{(1)}}\frac{Z^{({\hat s})}}{(Z_\delta)^2}.
\ee

To obtain the normalized 4-point function we write
\bea
\langle\sigma_2(0)\sigma_2(1)\sigma_2(w)\sigma_2(\infty)\rangle&\equiv&
\lim_{|z_\infty|\rightarrow\infty}|z_\infty|^{4\Delta_2}
\langle\sigma_2(0)\sigma_2(1)\sigma_2(w)\sigma_2(z_\infty)\rangle
\nonumber \\
&=&\lim_{|z_\infty|\rightarrow\infty}|z_\infty|^{4\Delta_2}
{\langle\sigma^\epsilon_2(0)\sigma^\epsilon_2(1)\sigma^\epsilon_2(w
)\sigma^{\tilde\delta}_2(z_\infty)\rangle\over
\langle \sigma^\epsilon_2(0)\sigma^\epsilon_2(1)\rangle^2}\nonumber\\
&=&2^{-2/3}|w(1-w)|^{-{1\over 12}}Z_\tau
\label{nfive}
\eea
Here we have used the fact that in this case the partition function
$Z^{(\hat s)}$ in (\ref{thir}) is that on the flat torus with
modular parameter $\tau$ given through (\ref{ntwo}).

Since the group $S^2$ equals the group $Z_2$, we can compare
(\ref{nfive}) with the 4-point function obtained
for $\sigma_2$ operators for the $Z_2$ orbifold in \cite{dixon}.
(\ref{nfive}) agrees with (4.13)  of \cite{dixon}
for the case of a noncompact  boson field and with (4.16) for the
compact boson field.

One observes that if the fields $X^i$ are noncompact bosons, then as
$w\rightarrow 1$ we find a factor $\log (w-1)$ in the OPE in addition 
to the expected
power $(w-1)^{1/4}$. We  suggest the following interpretation of this 
logarithm.
There is a continuous family of momentum modes for the noncompact boson, with
energy going to zero. If we do not orbifold the target space, then momentum
conservation allows only a definite momentum mode to appear in the OPE of two
fields. But the orbifolding destroys the translation invariance in $X^1-X^2$,
and nonzero momentum modes can be exchanged between sets of operators where 
each
set does not carry any net momentum charge. The exchange of such 
modes (with dimensions accumulating to zero)  between the pair 
$\sigma_2(w)\sigma_2(1)$ and the pair
$\sigma_2(0)\sigma_2(\infty)$ gives rise to the logarithm. Of course when the
boson is compact, this logarithm disappears, as can be verified from
(\ref{nfive}) or the equivalent results in \cite{dixon}.

\section{Discussion}
\renewcommand{\theequation}{8.\arabic{equation}}
\setcounter{equation}{0}

The motivation for our study of correlation functions of symmetric
orbifolds was the fact that the dual of the $AdS_3\times
S^3\times M$ spacetime (which arises in black hole studies) is the
CFT arising from the low energy limit of the D1-D5
system, and the D1-D5 system is believed to be a deformation of an
orbifold CFT (with the undeformed orbifold as a special
point in moduli space).  To study this duality we must really study
the supersymmetric orbifold theory, while in this
paper we have just studied the bosonic theory. It turns out however
that the supersymmetric orbifold can be studied
with only a small extension of what we have done here.  Following
\cite{jevicki} we can bosonize the fermions. Then if we go to the
covering space $\Sigma$ near the insertion of twist operator then we
find only the following difference from the
bosonic case -- at the location of the twist operator we do not have
the identity  operator, but instead a `charge operator' of
the form
$P(\partial X_a, \partial^2 X_a, \dots)e^{i\sum k_a X_a}$.
Here $X_a$ are the  bosons that arise from bosonizing the
complex fermions, and $P$ is a polynomial expression in its
arguments.  It is easy to compute the correlation function of
these charge operators on the covering space
$\Sigma$, and then we have the twist correlation functions for the
supersymmetric theory. We will present this
calculation elsewhere, but here we note that many properties of
interest for the supersymmetric correlation functions can
be already seen from the bosonic analysis that we have done here. In
this section we recall the features of the $AdS/CFT$
duality map and analyse some properties of the 3-point functions in the CFT.

\subsection{`Universality' of the correlation functions}

We have mentioned before that while we have  discsussed the
orbifold theory $R^N/S_N$  (where the coordinate of $R$ gave  $X$,  a
real scalar field), we could
replace the CFT of $X$ by any other CFT of our choice, and
the calculations performed here would remain essentially the same.
When the covering surface $\Sigma$ had genus zero,
the results depend only on the value of
$c$, and thus if we had
$(T^4)^N/S_N$ theory or  a $(K_3)^N/S_N$ theory, then we would simply
choose $c=4$ in  the Liouville action
(\ref{fourt})  (instead of $c=1$).  If $\Sigma$ had $g=1$, then we
would need to put in the partition function (of a single copy)
of $T^4$ or $K_3$ for the value of $Z^{(\hat s)}$ in (\ref{thir}).
But apart from the value of $c$ and the value of partition
functions on $\Sigma$  there  is no change in the calculation. Thus
in particular the 3-point functions that we have
computed at genus zero are universal in the sense that if we take
them to the power $c$ then we get the 3-point functions
for any CFT of the form $M^N/S_N$ with the CFT on $M$ having cenral charge $c$.

There is a small change in the calculation when we consider the
supersymmetric case. The fermions
from different copies of $M$ anticommute, and the twist operators 
carry a representation of the $R$
symmetry.  As a consequence  the dimensions of the twist
operators are not given by (\ref{teight}), but
for  the supersymmetric theory based on  $M=T^4$
are given by  ${1\over 2}(n-1)$. However as mentioned above, our
analysis can be
extended with small modifications to such theories as well.

Note that our method does not work if we have an orbifold group other
than $S_N$. Thus for example if we had a $Z_N$
orbifold of a complex boson \cite{dixon}, then we could go to the  covering
space over a twist operator $\sigma_n$, but  not write the
CFT in terms of an unconstrained field on this covering space. The
reason is that we have $n$ sheets  or more of the cover
over any point in the base space, but the central charge of the
theory is just $2$, and so we cannot attribute one scalar
field to each sheet of the cover.  Thus our method, and its
associated universalities, are special to $S_N$ orbifolds, where
a twist operator  just permutes copies of a given CFT but does not
exploit any special symmetry of the CFT itself.

\subsection{The genus expansion and the fusion rules of WZW models}

We have studied the orbifold CFT on the plane, but found that the
correlation functions can be organized in a genus
expansion, arising from the genus of the covering surface $\Sigma$.
In the large $N$ limit the contribution of a
higher genus surface goes like
$1/N^{g+{1\over 2}}$.  This situation is similar to that in the
Yang-Mills theory
that is dual to $AdS_5 \times S^5$. The Yang-Mills theory
has correlation functions that can be expanded in a genus expansion,
with higher genus surfaces supressed by $1/N^g$. In
the Yang-Mills theory the genus expansion has its origins in the
structure of Feynman diagrams for fields carring two
indices (the `double line representation' of gauge bosons). In our
case we have quite a different origin for the genus
expansion.  In the case of $AdS_5\times S^5$ it is believed that the genus
expansion of the dual Yang-Mills theory is related to the
genus expansion of  string theory  on this spacetime, though the
precise relationship is not clear. It would be
interesting if the genus expansion we have for the
$D1-D5$ CFT would be related to the genus expansion of the string
theory on $AdS_3\times S^3\times M$.

In this context we observe the following relation.  It was argued in
\cite{martinec}  that  the orbifold CFT $M^N/S_N$  indeed
corresponds to a point in the D1-D5 system moduli space. Further, at
this point we have the number of 1-branes
($n_1$) and of 5-branes ($n_5$)  given by $n_5=1, N=n_1n_5=n_1$. The
dual string theory is in general an $SU(2)$
Wess-Zumino-Witten (WZW) model \cite{giveon}, though at the orbifold
point of the
CFT this string theory is
complicated to analyze.  The  twist operators $\sigma_n$, $n=1\dots N$
of the CFT ($\sigma_1=$Identity) correspond to WZW
primaries  with $j=(n-1)/2, ~0\le j\le{N-1\over 2}$.  Since in a
usual WZW model we have $0\le j\le k/2$, we set $k=N-1$.

The fusion rules for the WZW model, which give the 3-point functions
of the string theory on the sphere (tree level) are
as follows. The spins $j$ follow the rules for spin addition in
$SU(2)$, except that there is also a `truncation from above'
\bea
(j_1, j_2)&\rightarrow & j_3\nonumber\\
|j_1-j_2|&\le& j_3~\le~ |j_1+j_2|, ~~~~~~~j_1+j_2+j_3~\le ~k
\label{oone}
\eea

      Now  consider the 3-point function in the orbifold CFT, for the
case where the genus of the covering surface $\Sigma$ is
$g=0$ . The ramification order of
$\Sigma$  at the insertion of $\sigma_{n_i}$ is $r_i=(n_i-1)=2j_i$.
The rules in (\ref{thone}), (\ref{thtwo}), (\ref{ththree})
translate to $|j_1-j_2|\le j_3~\le~ |j_1+j_2|$. Further, the
number of sheets $s$ is bounded as $s\le N$. Then the  relation
(\ref{tnine}) gives
\be
\sum_i {r_i\over 2}=g-1+s\le -1+N~~\rightarrow ~~ j_1+j_2+j_3\le k
\label{otwo}
\ee
While (\ref{otwo}) is a relation for the bosonic orbifold theory, we
expect an essentially similar relation for the
supersymmetric case. Thus we observe a similarity between the $g=0$
3-point functions of the WZW model (\ref{oone}) and
of the CFT (\ref{otwo}).

At genus $g=1$ however, we find that any three spins $j_1, j_2, j_3$
can give a nonzero 3-point function in the string
theory. In the orbifold CFT, however, we get only a slight relaxation
of the rule (\ref{otwo}): we get $j_1+j_2+j_3\le k+1$.
Roughly speaking we can reproduce this rule in the string theory if
we require that in the string theory one loop diagram
there be a way to draw the lines such that only spins $j\le 1/2$ be
allowed to circulate in the loop. Of course we are outside
the domain of any good perturbation expansion at this point, since if
the spins are of order $k$ then there is no small
parameter in the theory to expand in, and thus there is no
requirement that there be an exact relation between the rules
in a WZW string theory and the rules in the orbifold CFT.

We note that in \cite{jevicki} the 3-point functions of chiral primaries
that were studied had `one overlap' in their indices.
This corresponds to $j_1+j_2=j_3$ in the above fusion rules, and
since for the supersymmetric case the dimension is linear in
the charge, we also have $\Delta_1+\Delta_2=\Delta_3$.  This
corresponds to the case of `extremal' correlation functions in
the language of \cite{dhoker}. In \cite{jevicki}  the 3-point
correlators for this
special case were found by an elegant recursion relation,
which arises from the fact that there is no singularity in the OPE,
and thus the duality relation of conformal blocks becomes
a `chiral ring' type of associativity law among the fusion
coefficients. It is not clear however how to extend this method to
the non-extremal case, and one motivation for the present work was to
develop a scheme to compute the correlators for
$j_1+j_2<j_3$, which corresponds to more than one overlap.  In the
case of one overlap we have extended our calculations
to the supersymmetric case, and found results in agreement with \cite{jevicki}.

\subsection{ 3-point couplings and the stringy exclusion principle.}

In the $AdS_5\times S^5$ case the 3-point couplings of supergravity
agree with the large $N$ limit of the 3-point functions
in the free Yang-Mills theory; thus there is a nonrenormalization of
this correlator as the coupling $g$ is varied. It is not
clear if a similar result holds for the $AdS_3\times S^3\times M$
case, and even less clear what nonrenormalization
theorems hold at finite $N$.  But it is nevertheless interesting to
ask how the correlators in the orbifold CFT behave as we
go from infinite $N$ to finite $N$, and in particular what happens as
we approach the limits of the stringy exclusion principle.

Thus we examine the ratio
\be
R(m,n,q;\bar N)~\equiv ~ {\sqrt{\bar N}\langle O_nO_mO_q\rangle_{\bar N}\over
\lim_{N\rightarrow\infty}\sqrt{N}\langle O_nO_mO_q\rangle_{N}}
\label{othree}
\ee
where the subscripts on the correlator give the value of $N$.  We
have rescaled the correlators by $\sqrt{N}$  to obtain
the effective coupling of the 3-point function; the correlator itself
goes as $1/\sqrt{N}$. For
$n,m,q<<\bar N$ we expect
$R\approx 1$, while as
$n,m,q$ become order $\bar N$ we expect that $R$ will fall to zero.
We take the case of the 3-point function with single
overlap, and further set $m=n$. Then we have $q=2n-1$, and we write
\be
R(n,n,2n-1;\bar N)\equiv R(n;\bar N)
\label{FactorToPlot}
\ee
It is easy to see that for the case of single overlap the
correlators
$\langle\sigma_n\sigma_m\sigma_{m+n-1}\rangle$ can get a contribution
only from surfaces $\Sigma$ with $g=0$,  for
which case we have done a complete calculation of the correlator and
its combinatorics. Note further that in the ratio
(\ref{othree}) the actual value of
$\langle\sigma_n\sigma_m\sigma_{m+n-1}\rangle$ will cancel, and the
value of $R$ will be
determined by combinatorial factors. These factors are expected to
be the same for the bosonic and supersymmetric
cases.

In the figure we plot  $R(n;\bar N)$ versus $n$ (for $\bar
N=1000$).  We see that $R$ drops
significantly after $n$ exceeds $\sim \sqrt{\bar N}$.
This effect can be traced to the fact that the number of ways to
select $s$ ordered indices from $\bar N$ indices is
\bea
\bar N(\bar
N-1)&\dots& (\bar N-s+1)=\bar N^s(1-{1\over \bar N})(1-{2\over \bar
N})\dots (1-{s-1\over \bar N})\nonumber \\
&\approx&
\bar N^s(1-{1\over
\bar N}\sum_{j=1}^{s-1} j)=\bar N^s(1-{1\over
\bar N}{s(s-1)\over 2})
\label{ofour}
\eea

\begin{figure}
\epsfxsize=5in \epsffile{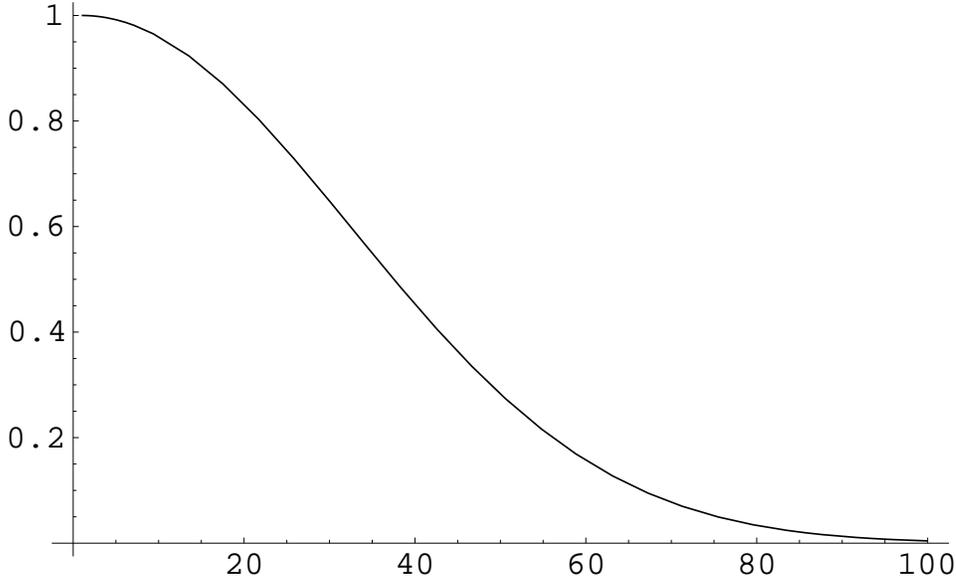}
\caption{
The factor (\ref{FactorToPlot}) as function of $n$ for
$\bar N=1000$.
\label{figure}}
\end{figure}

If the CFT 3-point function is  indeed not renormalized for finite
$N$, then the above result has interesting
implications.  The coupling between three gravitons would then be a constant
for low energies ($n<<\bar N$) but would drop rapidly for
high energies. Thus the behavior of high frequency modes would not
follow a naive  `equivalence principle'.

This issue may be relevant to  Hawking's derivation of black hole
radiation, where we need to make a change
of coordinates to study the high frequency modes near the horizon.
For these modes to evolve as used in the
derivation, we use implicitly the naive value of  the following
cubic coupling: that  of a low energy graviton (representing the
attraction  of the hole) and two high energy quanta
(representing the high energy mode emerging from the horizon, getting
redshifted by the attraction of the hole). If this
coupling differs from the one expected from naive gravitational
physics,   then the semiclassical
derivation of Hawking radiation may require modification,  with
corresponding implications for the
information paradox.

\subsection{Conclusion}

It would be important to pursue further the study of the
supersymmetric case, and to compare with
the dual superstring theory.  The subset of correlators computed in
\cite{jevicki} was compared to
supergravity in \cite{mihailescu},  but it was a little unclear how
closely the two calculations agreed. A better
picture may emerge when we look at the complete set of correlators of
the supersymmetric side, which is
possible to do by extending our computation here to include the
R-charges carried by the twist operators in
the supersymmetric case.

   It was argued recently in
\cite{farey} that the CFT of the D1-D5 system exhibits a duality to a
set of spacetimes,  of which the AdS space
is only one member.  If the 3-point functions are protected against
coupling changes then we should see a
reflection of this fact in correlators at the orbifold point.

\section*{Acknowledgements}

We are grateful to A. Jevicki, M. Mihailescu, S. Ramgoolam, and S.
Frolov for patiently explaining their results to us, and to  L.
Rastelli for extensive discussions in the early phase of this work.
We also benefited greatly from discussions with  S. Das, E.
D'Hoker,  C. Imbimbo,  F. Larsen,  E. Martinec,  S. Mukhi,  S. Sethi,
and  S.T. Yau.

\end{document}